\documentclass[%
 reprint,
 amsmath,amssymb,
 aps,
]{revtex4-2}

\usepackage{color}
\usepackage{graphicx}
\usepackage{dcolumn}
\usepackage{bm}

\usepackage{amsmath}
\usepackage{amssymb}
\usepackage{mathtools}

\newcommand{\sign}{\text{sign}}

\begin{document}

\preprint{APS/123-QED}

\title{Driven particle dispersion in narrow disordered racetracks}

\author{Federico El\'ias}
\email{federico.elias@cab.cnea.gov.ar}
\affiliation{Centro At\'omico Bariloche and Instituto Balseiro,
CNEA, CONICET and Universidad Nacional de Cuyo, 8400 Bariloche, Argentina}
\affiliation{}

\author{Alejandro B. Kolton}
\email{alejandro.kolton@ib.edu.ar}
\affiliation{Centro At\'omico Bariloche and Instituto Balseiro,
CNEA, CONICET and Universidad Nacional de Cuyo, 8400 Bariloche, Argentina}

\date{\today}

\begin{abstract}
We study the disorder-induced deterministic dispersion of particles uniformly driven in an array of narrow tracks. For different
toy models with quenched disorder we obtain exact analytical expressions for the steady-state mean velocity $v$ and the dispersion constant $D$ for any driving force $f$ above the depinning threshold. For short-range correlated pinning forces we find that at large drives $D\sim 1/v$ for random-field type of disorder while $D \sim 1/v^3$ for the random-bond type. We show numerically that these results are robust: the same scaling holds for models of massive damped particles, soft particles,  particles in quasi-one dimensional or two dimensional tracks, and for a model of a magnetic domain wall with two degrees of freedom driven either by electrical current or magnetic field. Crossover and finite temperature effects are  discussed. The universal features we identify may be relevant for describing the fluctuating dynamics of stable localized objects such as solitons, superconducting vortices, magnetic domain walls and skyrmions, and colloids driven in quasi one-dimensional track arrays. In particular, the drive dependence of $D$ appears as a sensitive tool for characterizing and assessing the nature of disorder in the host materials.
\end{abstract}

\maketitle

\section{Introduction}

A precise control of the motion in narrow tracks of stable localized excitations is desirable to realize different devices of current interest in applied physics research.
Paradigmatic examples are magnetic domain walls ~\cite{Parkin2008,Emori2013,Blasing2020} or skyrmions \cite{Kiselev2011,Schulz2012,Fert2013,Reichhardt2021} moving in an array of magnetic nanowires. 
Furthermore, such kind of devices may be also of interest for other driven objects embedded in different materials, such as ferroelectric domain walls~\cite{Catalan2012,Paruch2013}, vortices in superconductors~\cite{Kwok2016}, colloids~\cite{Lowen2008,Volpe2014,Bechinger2016}, general brownian particles~\cite{hanggi2009}, as well as kinks and solitons~\cite{Martinez2008,Laliena2020,Osorio2021,Osorio2022}.
Although each one may have a different internal structure we will generically refer to all these driven objects as ``particles''.

One of the interesting applications proposed for the above mentioned devices is the ``racetrack memory''~\cite{Parkin2008}, a novel type of non-volatile memories where particles are used as information carriers in an array of tracks. This proposal assumes the ability
to write, read and move the particles in the tracks. To achieve a competitive capacity and efficiency compared with existing memories, these tracks have been reduced to the nanoscale, making the consideration of (the usually unavoidable) quenched spatial heterogeneities in the host materials of prime importance. 
Indeed, spatial inhomogeneities may have disruptive effects, mainly metastability, a depinning transition, and a shaking effect on the sliding particles. As a consequence, the motion of the driven particles becomes more difficult to predict, specially in the ubiquitous case of quenched disorder. Moreover, in the case of track arrays, as each track has a different realization of the quenched disorder, different tracks may produce a different particle trajectories under an identical protocol. This is also a very important issue if we aim to synchronize the motion of many particles in many independent tracks, such as in a ``memory bus'' and then, for instance, perform a computation with the whole bunch of data. 
Although the effects of quenched disorder or quenched periodic potentials on driven particles have been extensively discussed in the literature, particularly focusing in the collective transport in two dimensions or higher, both on plastic ~\cite{Reichhardt2016,Reichhardt2021} or elastic flows~\cite{Wiese2022}, 
a detailed statistical study of the effect of quenched disorder on the somehow simpler yet experimentally relevant case of independent particles driven in an array made of quasi one-dimensional racetracks, is lacking.

In this paper we address the problem of independent particles driven in an array of disordered tracks with the aim to identify robust statistical features, independent of specific model details. To be concrete we focus in the specific ``obstacle race'' problem illustrated schematically in Fig.\ref{fig:dispersion}. At $t=0$ many particles start at $x=0$ in a track array, one per track, and are driven by the same constant force $f$. As tracks are independent realizations of an otherwise statistically identical quenched disorder with some assumed short-range correlations, at any given time $t>0$ particles disperse, even at zero temperature, provided the force exceeds the depinning threshold (assumed to be finite). We will be mainly interested in the mean velocity $v$ and in the dispersion constant $D$ around the mean forward motion of all the particles, as a function of $f$. Such race problem is similar, except for some subtle differences we discuss, to the well studied problem of dispersion of tracers in porous media. However, the results turn out to be radically different: while in the latter the dispersion constant $D$ in general increases with increasing the average flow~\cite{BouchaudGeorges1990}, in the problem we address here $D$ decreases with increasing the average flow $v$. We find, in particular, that $D$ has rather robust properties at large $f$, in the limit when $v\to f$, though quite sensitive to the nature of disorder. For the two physically important cases that we study here, the so-called Random-Bond (RB) and the Random-Field (RF) type, we find that $D\sim f^{-3}$ and $D\sim f^{-1}$ respectively. 
These results are obtained from different toy models where the full velocity-force characteristic and dispersion can be analytically obtained.
More complicated models, motivated in different physically relevant situations are solved numerically and shown to display the same dispersion beyond a velocity crossover.

The paper is organized as follows. In section \ref{sec:mechanicalmodel} we introduce the main properties of interest in relation to a simple mechanical model. Then, in section \ref{sec:exactresults} we introduce several models that can be solved analytically for the properties of interest, $v$ and $D$. These examples illustrate various effects and in particular suggest a universal behaviour of the dispersion constant $D$. In section \ref{sec:trapmodel} we show how these results can be connected to those for a stochastic trap model in the context of flow in porous media.
In order to test the robustness of the different analytical results, in section \ref{sec:numerics} we compare them with the numerical solution of various physically relevant models for which an exact solution is difficult to obtain. Finally, in section \ref{sec:discussion} we discuss the results and summarize our conclusions. 


\section{A mechanical toy model}
\label{sec:mechanicalmodel}
To be concrete let us consider a damped particle of mass $m$ driven by a constant uniform force $f$ in a one-dimensional space presenting quenched random forces at {\it zero temperature} (see section \ref{sec:langevin} for a discussion about temperature effects),
\begin{equation}
m \ddot x    = {F}({x})+ {f} - \eta \dot {x},
\end{equation}
where $\eta$ is a friction constant and ${F}(x)$ is a short-range correlated quenched random force field such that
\begin{eqnarray}
\overline{{F}({x})}&=&0,\\
\overline{{F}({x}){F}({x'})}&=& f_0^2 g(|{x}-{x'}|/d_0).
\label{eq:dampedmechmodel}
\end{eqnarray}
Here $d_0$ is a characteristic length, 
$f_0$ a characteristic force amplitude, and $g(u)$ a rapidly decaying function of unit range and $g(0)=1$. We will be particularly interested in the cases where $\int_x g(x)=0$, corresponding to the so-called ``Random-Bond'' case (RB), and the case where $\int_x g(x)>0$, corresponding to the so-called  ``Random-Field'' (RF) case. If we write $F(x)=-U'(x)$ (which is always possible to do in $d=1$ but not necessarily in $d>1$), and the potential $U(x)$ is bounded, then $F(x)$ is RB type. If otherwise the potential diffuses as $\langle [U(x)-U(x')]^2 \rangle \sim |x-x'|$ for long $|x-x'|$, then it is of the RF type. The great physical relevance of these two particular cases is better appreciated by noting that the RB type can be generated by forces derived from a bounded short-range correlated random potential while the RF type can be generated by a short-range correlated random force-field. 
In both cases we will also assume that $F(x)$ is bounded, so a finite critical depinning force $f_c = \max_x[-F(x)]$ exists, 
such that only for $f>f_c$ a steady-state motion is generated. 
\footnote{See Ref.\cite{LeDoussal2009}for a thorough and general study of the one particle depinning in different universality classes corresponding to the three extreme value statistics, Gumbel, Weibull, and Fréchet.} 

Without any loss of generality 
we can nondimensionalize the equation of motion Eq.(\ref{eq:dampedmechmodel}) by measuring distances in units of $d_0$, forces in units of $f_0$ and time in units of $\tau_0=\eta d_0 /f_0$, and mass in units of $m_0=\eta^2 d_0/f_0$. The derived velocity unit is therefore $v_0=d_0/\tau_0=f_0/\tau_0 \eta$.
Then, overriding notation for all nondimensionalized quantities we get 
\begin{equation}
m \ddot x    = {F}({x})+ {f} - \dot {x}.
\label{eq:equationwithinertia}
\end{equation}
For a given initial condition and disorder realization $F(x)$ the solution $x(t)$ of this equation is only parameterized by the nondimensional mass $m$. 
We will be mainly interested in the statistics of the time-dependent fluctuations induced by quenched disorder. 
Given the position $x(t)$ of a particle as a function of time $t$ for many particular realizations of $F(x)$, we will be mainly interested in the average velocity $v$ and the dispersion constant $D$ in the commoving frame at long times, as a function of the driving force $f$ and in the steady-state regime. Assuming $x(t=0)=0$, they are defined as
\begin{eqnarray}
v &=& \lim_{t\to \infty} \langle x(t)\rangle/t 
\label{eq:vdef} \\
D &=& \lim_{t\to \infty} \langle [x(t)-v t]^2 \rangle/t  
\label{eq:Ddef}
\end{eqnarray}
where the $\langle \dots \rangle$ denotes average over disorder realizations
\footnote{Note that we are thus interested here in the so-called ``annealead'' dispersion constant, different than the ``quenched'' one that describes the spreading of a packet in a single channel due to thermal noise or chaotic behaviour at zero temperature~\cite{LeDoussal1989}.}.
\begin{figure}[htp]
    \centering
\includegraphics[width=0.5\textwidth]{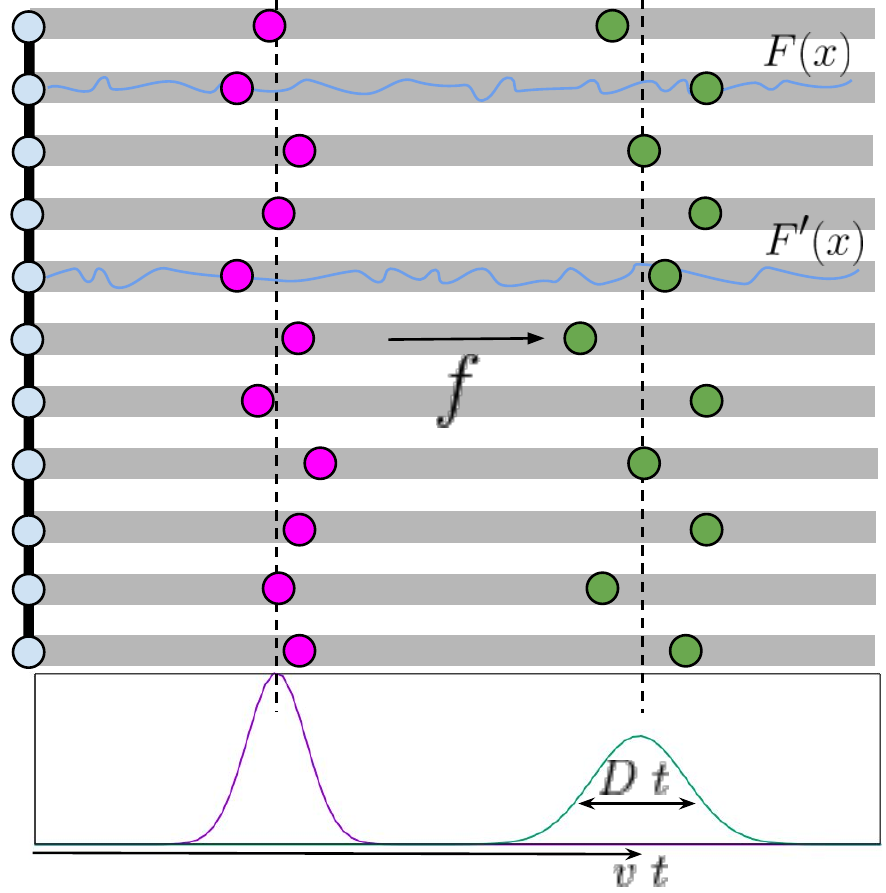}
    \caption{Schematics of the particle (filled circles) deterministic dispersion problem at finite force $f$ above the depinning threshold, induced by quenched disorder in an array of narrow disordered racetracks (grey bars). Tracks are different disorder realizations of statistically {\it identical} media (light-blue lines illustrate two force-fields $F(x)$ and $F'(x)$). Particles disperse around the average motion $vt$ in the the right direction, with $v$ the disorder-averaged or track-averaged velocity. The ``obstacle race'' starts at $t=0$, and for increasing times (indicated with different colors) the particle distribution is shown below. Dispersion is measured by the dispersion constant $D$. The dependence of $v$ and $D$ with $f$ contains useful and sensitive information about the type of disorder in the media. }
    \label{fig:dispersion}
\end{figure}
To give these definitions a concrete physical interpretation we can think of particles running on parallel racetracks, each one with its own realization $F(x)$ but identical statistical properties, with one particle per track (see Fig.\ref{fig:dispersion}). Then the average velocity $v$ is the center of mass velocity and $D t$ is the variance of their distribution around the center of mass, for a large collection of particles. 
We can imagine the channel array as a memory bus and the particles as carriers of information. In such a thought application $D$ and $v$ are relevant quantities to perform an efficient parallel computation. If particles are actually localized objects of a certain finite size, such as magnetic  domain walls, magnetic skyrmions,  solitons or colloidal particles, the zero temperature approximation may be well justified.

In addition to $v$ and $D$ we will be also interested in the steady-state force-dependent differential mobility, defined as
\begin{eqnarray}
 \mu = \frac{dv}{df}
\end{eqnarray}
allowing us to define an effective temperature from a generalized Einstein relation, 
as
\begin{eqnarray}
 T_{\rm eff} = \frac{D}{2\mu}.
\label{eq:Teffdefinition}
\end{eqnarray}
$T_{\rm eff}$ will in general depend on the driving force $f$, or on the velocity $v$.

\section{Exact results}
\label{sec:exactresults}
{\it Exact results for any $f$} can be readily obtained in the $m\to 0$ overdamped case 
\begin{equation}
    \dot x = F(x) + f.
    \label{eq:monomereq}
\end{equation}
when $F(x)$ is a piece-wise short-range correlated random function, such that the random-force $F(x)$ is constant in successive regular intervals (i.e. a random-steps force-field). If we then chose a bounded distribution for $F(x)$, a finite depinning threshold $f_c = \max[-F(x)]$, such that $v=0$ for $f<f_c$, is guaranteed to exist. Within this family of models we find (see appendix \ref{sec:appendixformulaD}), 
\begin{eqnarray}
v &=& 1/\langle \Delta t \rangle \label{eq:vformula}
\\
D &=& \frac{\langle \Delta t^2 \rangle-\langle \Delta t \rangle^2}{\langle \Delta t \rangle^3} 
\label{eq:Dformula}
\end{eqnarray}
where $\Delta t$ the time a particle spends on a given cell of size $\Delta x = 1$.
Therefore, the calculations of $v$ and $D$ reduce to single cell 
averages in the steady-state. 
Below we derive results for two particular cases, a Random-Field RF case and a Random-Bond RB case and later explore their 
robustness. 

\subsection{Box distribution of random forces}
\label{sec:boxdist}

Let us start with an instructive toy model with 
uniformly distributed bounded forces at each cell. 
\begin{figure}[htp]
    \centering
\includegraphics[width=0.5\textwidth]{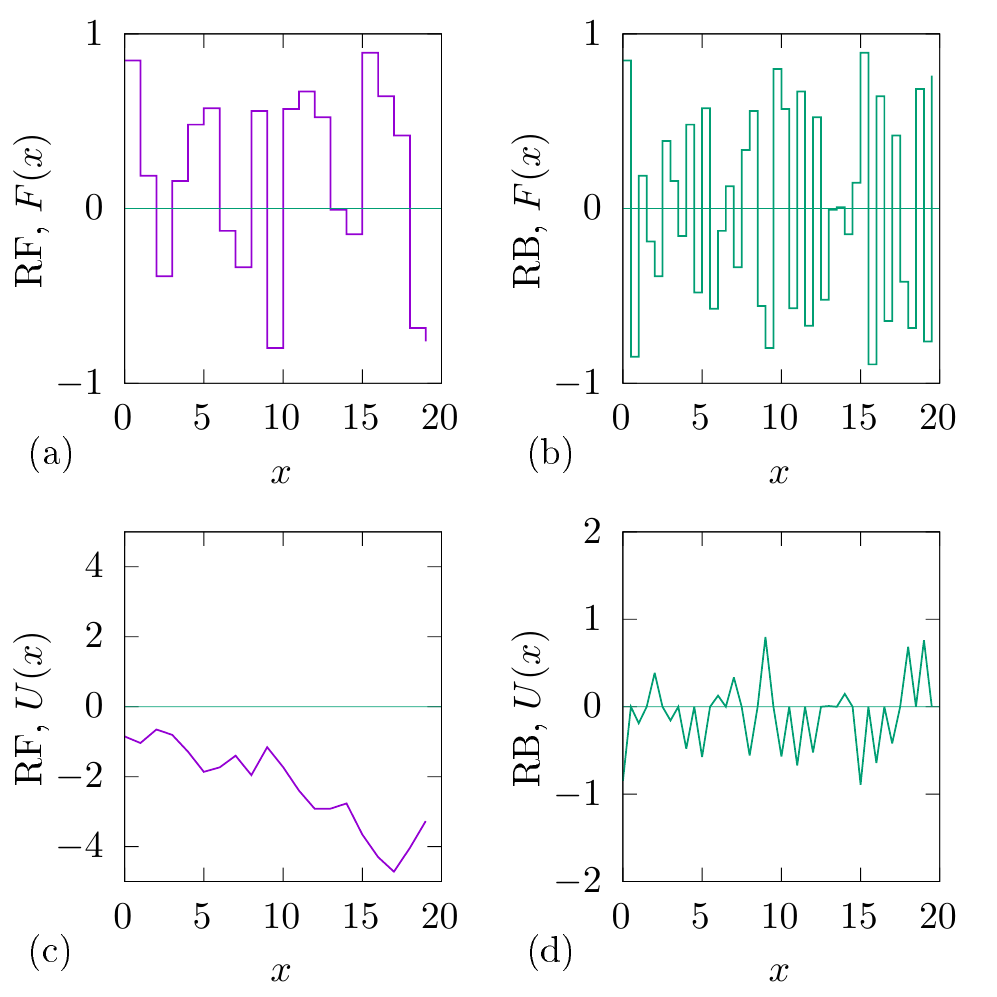}
    \caption{A piece-wise RF force field (a) and its potential (c). A piece-wise RB force field (b) and its potential (d). In both cases the depinning threshold is $f_c \to 1$ for large systems. Exact calculations show that both cases have the same $v$ but different $D$.
    }
    \label{fig:RFandRB}
\end{figure}
\subsubsection{RF case}
To construct a RF disorder such that $\int_y g(y)>0$ we  take
\begin{eqnarray}
F(x)=R_{[x]}
\end{eqnarray}
where $[\dots]$ denotes the integer part. Constant force intervals are hence of size $\Delta x=1$ and labeled with an integer $n$.
The $R_n$ are uniformly distributed 
random numbers in the interval $[-1,1]$
such that 
\begin{eqnarray}
\overline{R_n R_m} &=&\frac{\delta_{n,m}}{3} .
\end{eqnarray}
A sample of this RF force field, zoomed at short distances, is shown in Fig.\ref{fig:RFandRB}(a), and it corresponding potential 
$U(x)=-\int_x F(x)$ in in Fig.\ref{fig:RFandRB}(c).

From Eq.(\ref{eq:monomereq})
the time spent in the interval $n$
is 
\begin{eqnarray}
\Delta t_n = \frac{1}{R_n+f} , 
\end{eqnarray}
and hence
\begin{eqnarray}
\langle \Delta t\rangle &=&
\frac{1}{2} \int_{-1}^{1}
\frac{1}{R+f} \; dR 
= \frac{1}{2}\ln \left(\frac{f+1}{f-1}\right), \\
\langle \Delta t^2 \rangle &=&
\frac{1}{2} \int_{-1}^{1}
\frac{1}{(R+f)^2} \;  dR 
= \frac{1}{f^2-1}.\qquad 
\end{eqnarray}
The {\it exact} mean velocity for $f \geq 1$ is  
\begin{eqnarray}
v = 
\frac{2}{\ln \left(\frac{f+1}{f-1}\right)}, 
\label{eq:exactvuniform}
\end{eqnarray}
displaying a  depinning transition at $f=1$.
Approaching the threshold, $f\to 1+$, the velocity vanishes continuously as $v \sim -2/\ln (f-1)$. This model is thus peculiar from the standard critical phenomena point of view where we expect a power-law $v \sim (f-1)^\beta$, for an order parameter $v$ and a control parameter $f$ 
(see however section \ref{sec:sintheta}).
On the other hand, for $f\gg 1$, $v\simeq f$ as expected from the washing out of the pinning force at large velocities, where free flow is recovered. The velocity-force characteristics can be also inverted, yielding 
\begin{eqnarray}
 f = \frac{e^{2/v}+1}{e^{2/v}-1}.
\end{eqnarray}
The exact (differential) mobility for $f \geq 1$ is  
\begin{eqnarray}
\mu = 
\frac{4}{
(f^2-1) \ln^2 \left(\frac{f+1}{f-1}\right)},
\end{eqnarray}
such that $\mu \to \infty$ when $f\to 1$ (i.e. right at the depinning transition), and $\mu\to 1$ when $f\to \infty$ as expected.

Using Eq.(\ref{eq:Dformula}) the exact dispersion constant for $f \geq 1$ is
\begin{eqnarray}
 D = 
 \frac{8 \left(\frac{1}{f^2-1}-\frac{1}{4} \ln ^2\left(\frac{f+1}{f-1}\right)\right)}{\ln ^3\left(\frac{f+1}{f-1}\right)}
\label{eq:exactDRF}
\end{eqnarray}

and therefore, using the generalized Einstein relation we get the exact effective temperature
\begin{eqnarray}
 T_{\rm eff}=
  \frac{2  \left(1-\frac{f^2-1}{4} \ln ^2\left(\frac{f+1}{f-1}\right)\right)}{\ln \left(\frac{f+1}{f-1}\right)
  }
\end{eqnarray}
We are interested in the $f\gg 1$ fast-flow behaviour of $D$ and $T_{\tt eff}$. Expanding in powers of $1/f$ we first get
\begin{eqnarray}
v &\simeq& f - \frac{1}{3f} - \frac{4}{45 f^3} + \mathcal{O}(f^{-5}) 
\label{eq:largevexpansion}
\\
\mu &\simeq& 1 + \frac{1}{3f^2}+\mathcal{O}(f^{-4}),
\end{eqnarray}
and in particular for the dispersion and associated effective temperature we get
\begin{eqnarray}
D &\simeq& \frac{1}{3f} + \frac{7}{45 f^3 } + \mathcal{O}(f^{-5}) \nonumber \\
&\simeq& 
\frac{1}{3v} + \frac{2}{45 v^3} + \mathcal{O}(v^{-4}),\\
T_{\tt eff}&\simeq& \frac{1}{3f}+\frac{2}{45f^3}
+\mathcal{O}(f^{-5}).
\end{eqnarray}

We will particularly focus our attention on the large-velocity dominant term, 
which in this case is 
\begin{eqnarray}
 D\sim 1/f \sim 1/v,\;\;\; \text{RF}.
\label{eq:RFresult}
\end{eqnarray}
The above RF scaling can be obtained by first order perturbation theory in $f^{-1}$. At zero-th order 
Eq.(\ref{eq:monomereq}) implies $x \approx f t$, 
and then at first order $\dot{x}=f+F(f t)$. The random fluctuating force $F(ft)$ mimics a colored noise because $\langle F(ft) \rangle=0$, $\langle F(ft)F(ft') \rangle = g(f(t-t'))$, where $g(x)$ is short-ranged. Most importantly, because we are considering RF disorder, $\int g(x)>0$ and thus we can define a positive effective temperature $T_{\rm eff} \propto \int_{-\infty}^{\infty}  g(ft)dt \sim f^{-1}$. Hence, since $\mu\to 1$, $D\sim 1/f$. 

\subsubsection{RB case}
\label{sec:RBcaseUniform}
To model  RB disorder 
with $\int_y g(y)=0$, 
we define the random forces in Eq.(\ref{eq:monomereq}) as
\begin{eqnarray}
    F(x) = R_{[x]} \sign(x-[x]-1/2),
\end{eqnarray}
where as above the $R_n$ are iid random variables, uniformly distributed in  $[-1,1]$.
This choice splits the unit interval of each cell in two  parts of size $1/2$, such that the constant force in the second half is minus the force of the first half. 
A sample of this RF force field at short distances is shown in Fig.\ref{fig:RFandRB}(b), and it corresponding potential 
$U(x)=-\int_x F(x)$ in in Fig.\ref{fig:RFandRB}(d).
Evidently, this is a particular way to obtain the auto-correlation needed for a RB type of disorder. As we show below, the choice also has the advantage of leaving $v$ and $\mu$ invariant from $RF$ to $RB$.

By repeating the   procedure of the previous section we obtain   $\langle \Delta t \rangle$ as   for the RF case, and thus {\it identical} $v$ and $\mu$ as a function of $f$.
However, $\langle \Delta t^2 \rangle$ is different,
\begin{eqnarray}
\langle \Delta t^2 \rangle &=&
\frac{1}{2} \int_{-1}^{1}
\left[ \frac1{ 2(f+R)}  + \frac1{2 (f-R)} \right] ^2\; dR \\
&=& 
\frac{1}{4 f}
 \left[\frac{2f}{f^2-1}+\ln \left(\frac{f+1}{f-1}\right)\right].
\end{eqnarray}
The {\it exact} dispersion constant is
\begin{eqnarray}
D=
 \frac{2 \left(\frac{2}{f^2-1}-\log ^2\left(\frac{f+1}{f-1}\right)+\frac{\log \left(\frac{f+1}{f-1}\right)}{f}\right)}{\log ^3\left(\frac{f+1}{f-1}\right)}.
\label{eq:exactDRB}
\end{eqnarray}
$D$ thus diverges at the depinning threshold 
and at large driving forces it vanishes as 
\begin{eqnarray}
D \simeq T_{\tt eff} \simeq \frac{4}{45 f^3}   + \mathcal{O}(f^{-5}) 
\simeq \frac{4}{45 v^3} + \mathcal{O}(v^{-5}), 
\end{eqnarray}
which is clearly different than the RF case.
Then, the dominant term in this case is, 
\begin{eqnarray}
 D\sim 1/f^3 \sim 1/v^3,\;\;\; \text{RB}
\label{eq:RBresult}
\end{eqnarray}
faster than the RF case. 
It is worth noting that the $f^{-3}$ RB-scaling, at variance with the $f^{-1}$ RF-scaling, can not be obtained from first order perturbation in $f^{-1}$. Indeed, dispersion is (incorrectly) zero at first order, because for RB we have
$\int g(x)=0$.

\subsubsection{Summary}
\begin{figure}[htp]
    \centering
\includegraphics[width=0.5\textwidth]{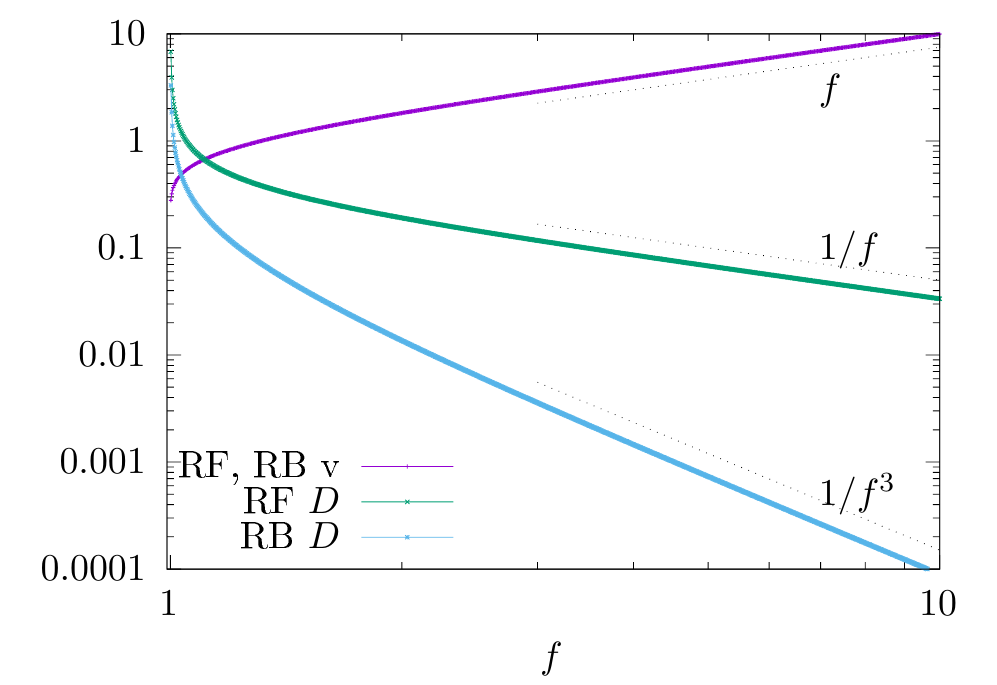}
    \caption{$v$ and $D$ vs $f$ for RF and RB force fields with a box-distribution. In this model, $v$ is invariant with respect to the RF or RB nature of the disorder. 
    At $f\to 1$ (depinning threshold) $v$ vanishes logarithmically, while $D$ diverges. At large forces, as indicated, $v \sim f$, while $D \sim 1/f$ and $D \sim 1/f^3$ for the RF and RB cases respectively, as indicated with dotted-lines.}
    \label{fig:boxdist}
\end{figure}

In Fig.\ref{fig:boxdist} we summarize the exact results for this model. 
We show that {\it identical} $v$ vs $f$ characteristics can be accompanied by different $D$ vs $f$ characteristics.
We show in particular that $D$ vanish differently at large $f$ in the RF and RB cases, thus becoming a sensitive tool to characterize the random-medium. It is worth noting that the decay of $D$ with $f$ depends crucially on the fact that the random-medium produce finite random forces whose effect becomes weaker at large velocities. A simple disordered model where this is not the case and $D$ increases with $f$ is discussed in the appendix \ref{sec:randomfriction}.

\subsection{``RB + $\epsilon$ RF'' case}
\label{sec:rb+rf}
We now discuss the case where disorder is not pure RB or RF but  where a RB disorder is perturbed by a RF disorder. To achieve this we split the $n$th unit interval in two equal parts, and put a uniformly distributed random force $R_n$ in the first half, followed by a random force $-R_n+\delta_n$ in the second half. We will consider that $\delta_n$ is a small random variable with $\langle \delta_n \rangle=0$ and $\langle \delta_n^2 \rangle=\epsilon^2 \ll 1$. 
We will also consider that $\langle R_m \delta_n\rangle=0$ so to ensure that $v$ remains invariant under the perturbation.

\begin{figure}[htp]
    \centering
\includegraphics[width=0.5\textwidth]{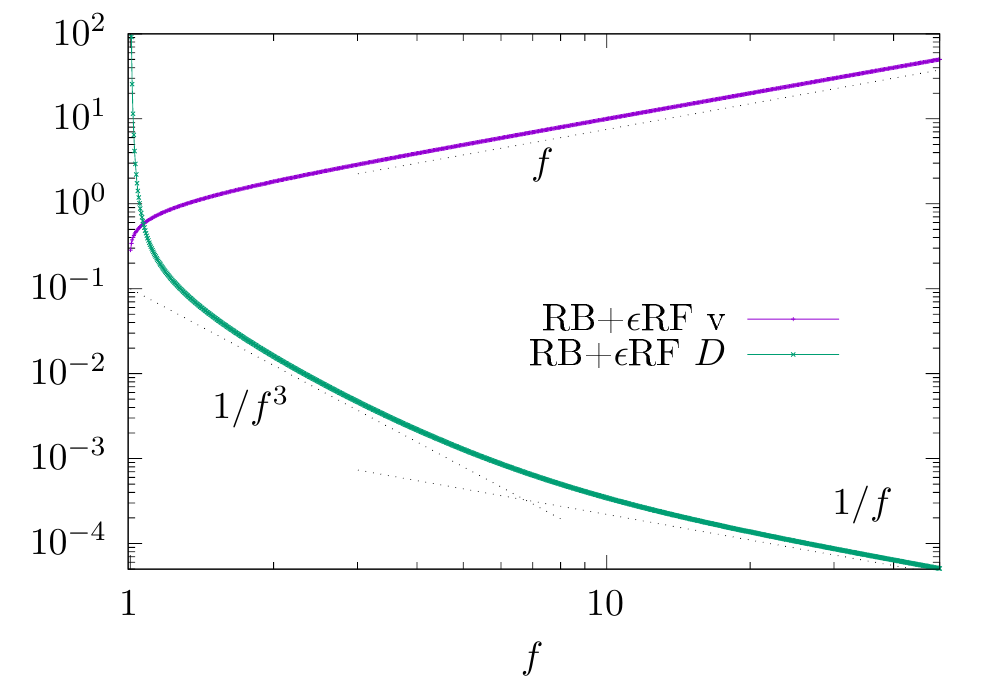}
    \caption{
    Crossover behaviour in a RB force field perturbed with a RF force field (both with a box-distribution of forces), for perturbing parameter $\epsilon=0.01$. While $v$ remains invariant under the small perturbation $D$ crossovers from $RB$ to $RF$ behaviour at large enough forces, as indicated with dotted-lines.
    }
    \label{fig:RBperturbRF}
\end{figure}

The $n$th residence time is
\begin{eqnarray}
    \Delta t_n &=& \frac{1}{2} \left( \frac{1}{f+R_n}+\frac{1}{f-R_n+\delta_n} \right) \nonumber \\
    &\approx& \Delta t_{RB} + \Delta t_{RF} 
    + \mathcal{O}(\epsilon^2)
\end{eqnarray}
where 
\begin{eqnarray}
 \Delta t_{RB} &=& \frac{f}{f^2-R_n^2}, \\
 \Delta t_{RF} &=& -\frac{\delta_n}{2(f-R_n)^2},
\end{eqnarray}
are the pure RB and RF random contributions.
Since $\langle R_m \delta_n\rangle=0$ 
we have
$\langle \Delta t_{RF} \rangle \propto  \langle \delta_n \rangle =0$. 
Therefore the RF perturbation does not alter the mean velocity, as $v^{-1}=\langle \Delta t_n \rangle=\langle \Delta t_{RB} \rangle$. 
Nevertheless,
\begin{equation}
    \langle \Delta t^2 \rangle  
    =  \langle \Delta t_{RB}^2 \rangle +
    \langle \Delta t_{RF}^2 \rangle 
\end{equation}
where in the last equality we have used $\langle \Delta t_{RB} \Delta t_{RF}\rangle \propto \langle \delta_n \rangle =0$. Since we already know 
$\langle \Delta t^2_{RB} \rangle$ from Sec.\ref{sec:RBcaseUniform} we have only to calculate $\langle \Delta t^2_{RF} \rangle$, which for uniformly distributed forces is simply
\begin{eqnarray}
\langle \Delta t^2_{RF} \rangle&=&
\langle \delta_n^2 \rangle \frac{1}{2}\int_{-1}^1 d R \left(\frac{1}{2(f-R)^2}\right)^2 
\nonumber \\
&=&
 \frac{\epsilon^2 \left(3 f^2+1\right)}{12 \left(f^2-1\right)^3}
\end{eqnarray}
Since
\begin{eqnarray}
D= \frac{
\langle \Delta t_{RB}^2 \rangle- \langle \Delta t^2 \rangle
+\langle \Delta t_{RF}^2 \rangle}{\langle \Delta t \rangle^3}
\end{eqnarray}
we obtain
\begin{eqnarray}
D= D_{RB} +  \frac{\langle \Delta t_{RF}^2\rangle}{\langle \Delta t\rangle^3} = D_{RB} + v^3 \langle \Delta t_{RF}^2\rangle.
\end{eqnarray}
To the leading term $\langle \Delta t_{RF}^2\rangle \approx 1/(4 f^4)$, $D_{RB}\approx 4/(45f^3)$ and $v \sim f$, we have
\begin{eqnarray}
 D \approx \frac{4}{45f^3} + \frac{\epsilon^2}{4f}. 
\end{eqnarray}
Therefore, the RF perturbation dominates at large enough velocities and $D \sim 1/f$. The crossover in this model roughly occurs at 
$f^* \sim \frac{4}{3\sqrt{5}\epsilon}$ if $\epsilon \ll 1$.
In Fig.\ref{fig:RBperturbRF} we summarize the exact results for this toy model.

The crossover just described may be important for instance for flat domain walls in media with RB disorder such as non-magnetic impurities, crystalline defects or rough borders, contaminated with a few magnetic impurities which produce RF pinning. On the other hand, it may be important for particles described by two or more coupled degrees of freedom, since in that case the force-field seen by the coordinate describing its position in the track is effective. 
More subtly, the crossover is important for the case of extended systems such as (non-flat) elastic interfaces but with pure RB microscopic disorder, since in that case the renormalized pinning force is not pure RB anymore~\cite{Wiese2022}.
At interface depinning indeed, the renormalized disorder becomes of a RF type and hence a merging of RF and RB into a unique RF depinning universality class emerges. However, as the velocity increases, such renormalization is expected to be weaker because the correlation length $l$ along the interface becomes smaller (near $f_c$ as $l \sim (f-f_c)^{-\nu}$), and then the renormalized pinning force tend to flow down to the RB microscopic disorder again. A toy model that beautifully illustrates this physics near the depinning threshold is the the particle quasistatically dragged in a random force field by a parabolic potential of curvature $m^2$~\cite{LeDoussal2009,terBurg2021,terBurg2022}. 
Then the effect of increasing $v$ can be related to the effect of increasing $m \sim 1/l$, and one can fully appreciate the flow from RB to RF by approaching the depinning threshold.
The results for the present toy model hence suggests that even the small RF part that remains from renormalization at large drives (or large $m$) should be still detectable beyond a crossover of $D$ as a function of $f$. 

\subsection{A non-uniform distribution of random forces}
\label{sec:sintheta}
In order to show the robustness of the previous results we now consider a simple model variant with the same scaling properties for $D$ at large velocities in the RF and RB case, but quite different critical behaviour near the depinning threshold. The particular variant proposed also displays much simpler exact analytical expressions for all quantities and could thus be used as a convenient toy model for other studies. 
In order to achieve this we change the probability distribution of random-forces in each cell such that
\begin{eqnarray}
 R_n = \sin \theta_n
\end{eqnarray}
with $\theta_n$ uniformly distributed in the $[-\pi,\pi)$ interval. This disorder may physically correspond to a particle coupled to the $x$ component of a random unit vector force in a plane.

\subsubsection{RF case}
As before, we compute the mean residence time in a cell
\begin{eqnarray}
\langle \Delta t \rangle =
\frac{1}{2\pi} \int_{-\pi}^{\pi}
\frac{d\theta}{\sin(\theta)+f} \; = \frac{1}{\sqrt{f^2-1}}, 
\end{eqnarray}
and hence the exact velocity and differential mobility are, respectively
\begin{eqnarray}
v = \sqrt{f^2-1}\\
\mu = \frac{f}{\sqrt{f^2-1}}.
\end{eqnarray}
Interestingly, this is exactly the mean velocity of the
$\dot x = f + \sin(x)$ model and
it has a finite and trivial critical exponent $v\sim (f-1)^{\beta}$, $\beta=1/2$, different than the previous logarithmic depinning case. Nevertheless, at variance with the particle in the periodic potential, where there is no dispersion at zero temperature, here we have a disorder-induced dispersion with respect to the disorder-averaged displacement. 
To see this we compute as before 
\begin{eqnarray}
\langle \Delta t^2 \rangle =
\frac{1}{2\pi} \int_{-\pi}^{\pi}
\frac{d\theta }{(\sin(\theta)+f)^2} \;
= \frac{f}{(f^2-1)^{3/2}}
\label{eq:nonUniformDt2}
\end{eqnarray}
Then, using Eq.(\ref{eq:Dformula}), we get the {\it exact} dispersion constant
\begin{eqnarray}
    D = f-\sqrt{f^2-1} 
    = -v + \sqrt{1+v^2}
\end{eqnarray}
which  {\it remains finite} as $f\to 1$ ($D\to 1$) and 
at large forces behaves as \begin{eqnarray}
    D  
    \approx 
    \frac{1}{2f}+\frac{1}{8 f^3} + \mathcal{O}(f^{-5})
    \approx 
    \frac{1}{2v}-\frac{1}{8v^3}+\mathcal{O}(v^{-5})
\end{eqnarray}
These results show that the RF dependence $D\sim 1/f$ at large velocities (Eq.(\ref{eq:RFresult})) is robust under the change of the step distribution we have made, in spite that the critical depinning behaviour is clearly different.

\subsubsection{RB case}
As before, we use the trick of dividing the unit interval in two correlated parts, 
\begin{eqnarray}
    F(x) = \sin \theta_{[x]} \sign(x-[x]-1/2).
\end{eqnarray}
where $\theta_n$ is uniformely distributed in $[\pi,\pi)$ as before. The convenience of this choice is, as before, that $v$ remains invariant.

The residence time in a single cell is then split in two contributions
\begin{eqnarray}
    \Delta t = \frac{1/2}{f+\sin(\theta)}+\frac{1/2}{f-\sin(\theta)}.
\end{eqnarray}
Therefore,
\begin{eqnarray}
\langle \Delta t \rangle &=&
\frac{1}{4\pi} \int_{-\pi}^{\pi}
\frac{d\theta}{\sin(\theta)+f} 
+\frac{1}{4\pi} \int_{-\pi}^{\pi}
\frac{d\theta}{-\sin(\theta)+f} 
\nonumber \\
&=& \frac{1}{\sqrt{f^2-1}},
\end{eqnarray}
{\it identical} to the RF case of the previous subsection, and hence $v$ and $\mu$ are also identical.
Nevertheless
\begin{eqnarray}
\langle \Delta t^2 \rangle =
\frac{2f^2-1}{2f(f^2-1)^{3/2}}
\end{eqnarray}
is different than the RF case, Eq.(\ref{eq:nonUniformDt2}). 
The {\it exact} dispersion constant is 
\begin{eqnarray}
 D=-\sqrt{f^2-1}+f-\frac{1}{2 f},
\end{eqnarray}
which {\it remains finite} as $f \to 1$ ($D\to 1/2$) and its large velocity expansion yields
\begin{eqnarray}
D &=& \frac{1}{8f^3 }+\mathcal{O}(f^{-5})
\end{eqnarray}
These results show that the RB dependence $D\sim 1/f^3 \sim 1/v^3$ at large velocities (Eq.(\ref{eq:RBresult})) is also robust under a change of the step distribution.

\begin{figure}[htp]
    \centering
\includegraphics[width=0.5\textwidth]{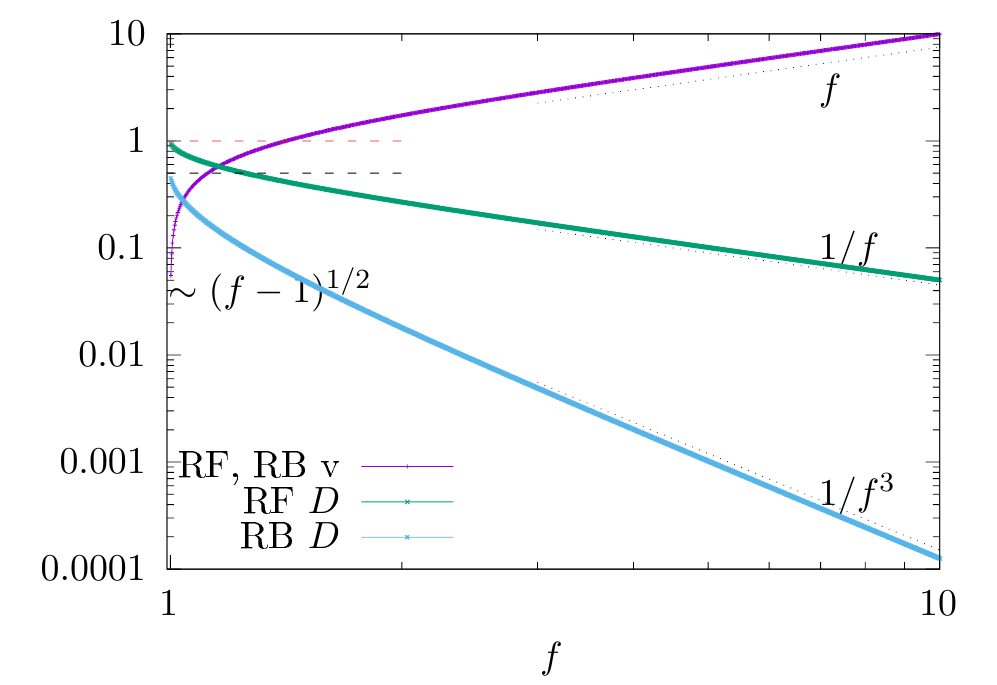}
    \caption{$v$ and $D$ vs $f$ for the RF and RB cases of the model with a  non-uniform force distribution. 
    At the depinning threshold $v\sim (f-1)^{1/2}$ and $D\to 1$ and $D\to 1/2$ (indicated with Red dashes lines) for RF and RB respectively, in sharp contrast with the box-distribution model.
    At large forces, we recover the robust $D\sim 1/f$ and $D\sim 1/f^3$ behaviours of the RF and RB cases respectively (indicated with dotted-lines).
    }
    \label{fig:sintheta}
\end{figure}

In Fig.\ref{fig:sintheta} we summarize the exact results for this model.
We note first that when $f\to 1+$, $v\sim (f-1)^{1/2}$,  in contrast with the logarithmic behaviour (Eq.(\ref{eq:exactvuniform})) obtained for the box distribution in the previous sections. Moreover, we find that $D \to 1$ (RF) and $D \to 0.5$ (RB), also in sharp contrast with the divergent dispersion constants in the same cases for the box distribution, Eqs.(\ref{eq:exactDRF}) and (\ref{eq:exactDRB}) 
respectively. 
Nevertheless, the (RF) $1/f$ and (RB) $1/f^3$ asymptotics of $D$ at large $f$ remains a robust feature. 

\subsection{Connection with a trap model for dispersion}
\label{sec:trapmodel}
Dispersion under an average flow is an old problem, relevant for the physics of porous media where the dispersion constant $D$ measures the spread of an injected packet of tracer particles. 
In this context a pedagogical toy model proposed by Bouchaud and Georges~\cite{BouchaudGeorges1990} allows to understand the basic physics and to estimate $D$. In that model the medium is idealized as one dimensional and made of a ``backbone'' along which the particle is convected with velocity $V$ at regularly spaced positions (separated by a distance $\xi$). With probability $p$ the particle can leave the backbone during a random waiting time $\tau$. Importantly, the particle can not go backwards. 
\begin{figure}[htp]
    \centering
\includegraphics[width=0.5\textwidth]{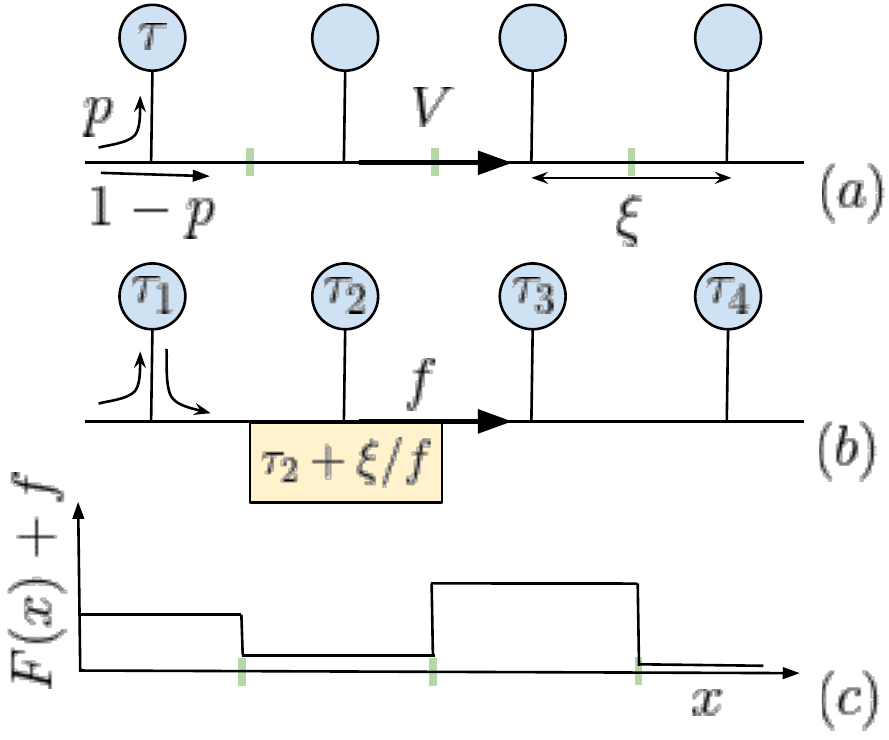}
    \caption{Adapting the Bouchaud and Georges trap model (a) to the quenched-force field toy model above the depinning threshold, at zero temperature (c). To convert it in (b) we consider that that free flow is $V=f$, and that the effective trapping occurs with probability one with waiting times that are random but quenched, each one given by the ``excess residence time'' in a cell $\tau_n \equiv \frac{\xi}{F_n+f}-\frac{\xi}{f}$, where $n\equiv [x]$ is the cell number.  }
    \label{fig:bouchaud}
\end{figure}
Bouchaud and Georges showed that 
\begin{eqnarray}
U^{-1}&=&V^{-1}+p\langle \tau \rangle/\xi 
\label{eq:bouchaudgeorgesU}
\\
D_\parallel &=& (pU^3/2\xi)(\langle \tau^2 \rangle - p \langle \tau \rangle^2).
\label{eq:bouchaudgeorgesD}
\end{eqnarray}
where $U$ is the average velocity of the particle in the medium and $D_\parallel$ denotes the longitudinal dispersion constant. These results show that for a given drive $V$, $D_\parallel$  {\it grows} with $V$, in good qualitative agreement with some porous media experiments. It is interesting to note that at large velocities this model predicts $D_\parallel \sim V^3$, in sharp contrast with our results for uncorrelated random media, $D\sim 1/f^3 \sim 1/v^3$ (RB) and $D\sim 1/f \sim 1/v$ (RF). 
In the following we explain such difference and show how to reconcile the results of Eqs.(\ref{eq:bouchaudgeorgesU}) and (\ref{eq:bouchaudgeorgesD})  with the predictions of our dispersion model. 

To explain the differences between the Bouchaud-Georges model predictions and the ones we presented in the previous sections we start by noting that the average in Eq.(\ref{eq:bouchaudgeorgesD}) is over different stochastic trajectories in one track, while in
Eq.(\ref{eq:Dformula}) the average is over disorder realizations (see Fig.\ref{fig:bouchaud}). The former is the so-called ``quenched'' dispersion constant while the later is the so-called ``annealed'' dispersion constant. We show below that this does not prevent however to find a concrete connection between the two mathematical approaches.
We also note that in the problem we are interested in, the delay $\langle \Delta t \rangle$ in each cell is not given by a fixed distribution, but by a drive dependent distribution. In particular, the larger $f$, the closer is $\langle \Delta t \rangle$ to $1/f$. 
Hence, to make a concrete connection with the overdamped mechanical model we replace the trap in the $n$th cell of size $\xi=1$ by a quenched random force field
and redefine $\tau$ as the {\it excess time} with respect to the free motion time in the cell $=1/f$. 
Note that replacing $\tau$ by quenched random values is not problematic because the particle never moves backward. That is, 
\begin{eqnarray}
\tau \equiv \Delta t - \frac{1}{f},  
\end{eqnarray}
and we also make the identification $V \equiv  f$ and $U \equiv v = 1/\langle \Delta t \rangle$. We can also take $p=1$ since the non-trapping event can be absorbed in the distribution of $\tau$, corresponding to $F_n=0$. With such analogy, noting that $2 D_\parallel \equiv D$ (because of the convention we have used to define $D$, see Eq.(\ref{eq:Ddef})), we finally get 
\begin{eqnarray}
v^{-1}&=& V^{-1}+ \langle \tau \rangle = {\langle \Delta t\rangle} \\
D &=& U^3 (\langle \tau^2 \rangle - \langle \tau \rangle^2)
= \frac{\langle \Delta t^2 \rangle - \langle \Delta t \rangle^2}{\langle \Delta t \rangle^3}
\end{eqnarray}
which is identical to our expressions, see Eq.(\ref{eq:Dformula}).

The later adaptation and reinterpretation of the different terms shows that our results for the mechanical models ultimately reduce to compute a few statistical properties (first and second moments) of the excess waiting time in a trap equivalent model.
Interestingly, the mechanical model yields always a {\it decreasing} $D$ with increasing $v$, and the microscopic nature of the short-correlated disorder is important to determine how exactly $D$ vanishes at large velocities.

\section{Numerical Results for different models}
\label{sec:numerics}
In order to assess the universality of some of the analytical predictions we have made in the previous sections for simple one-dimensional toy models, in this section we compare them with more complicated physically relevant disordered models, for which an exact solution is not available, but precise numerical simulations can be performed. 
We consider overdamped and damped models, the effect of temperature and the effect of transverse fluctuations in wires with a finite or infinite width. 
We also consider the dispersion of a non-rigid particle and a micromagnetic model of a magnetic domain wall driven by an external electrical current and/or applied magnetic field.

Solving all the proposed models reduces to integrate a system of ordinary non-linear differential equations of at most two variables, with disorder. In all cases we solve them by standard integration techniques. We use between $8192$ to $32768$ independent disorder realizations (or tracks) to average the properties of interest, and run long-time simulations so to capture the diffusive behaviour beyond any finite time crossover.

\subsection{Inertia: massive damped particles}
Embedded soft matter systems, such as magnetic domain walls, magnetic skyrmions, or superconducting vortices behave effectively as overdamped driven objects and inertia can be ignored. Inertia may be important however for other experimental realizations which could be modeled by a driven massive damped particle or soliton in a quenched random field. Inertia opens, on the other hand, a new phenomenology because the state space is doubled compared to the overdamped case. In our particular one-dimensional case of Eq.(\ref{eq:equationwithinertia}), the state of the damped particle must be described by two variables instead of one, and it is a priori not evident whether the universal results we obtain for $D$ in the RF and RB cases remain valid in this case. Since to the best of our knowledge no analytical solutions for the dispersion properties are known for Eq.(\ref{eq:equationwithinertia}) for finite $m$, here we solve it numerically.

For the simulations we solve the nondimensional Eq.(\ref{eq:equationwithinertia}) using the same piece-wise $F(x)$ considered in sections \ref{sec:boxdist} for the RB and RF cases. From  Eq.(\ref{eq:equationwithinertia}) 
we note that $v \approx f$ in the large $f$ limit, because the average acceleration must be zero in presence of damping. We will be particularly interest in the fluctuations when $f-v \ll f$.

\subsubsection{RF case}
\begin{figure}[htp]
    \centering
     \includegraphics[width=\columnwidth]{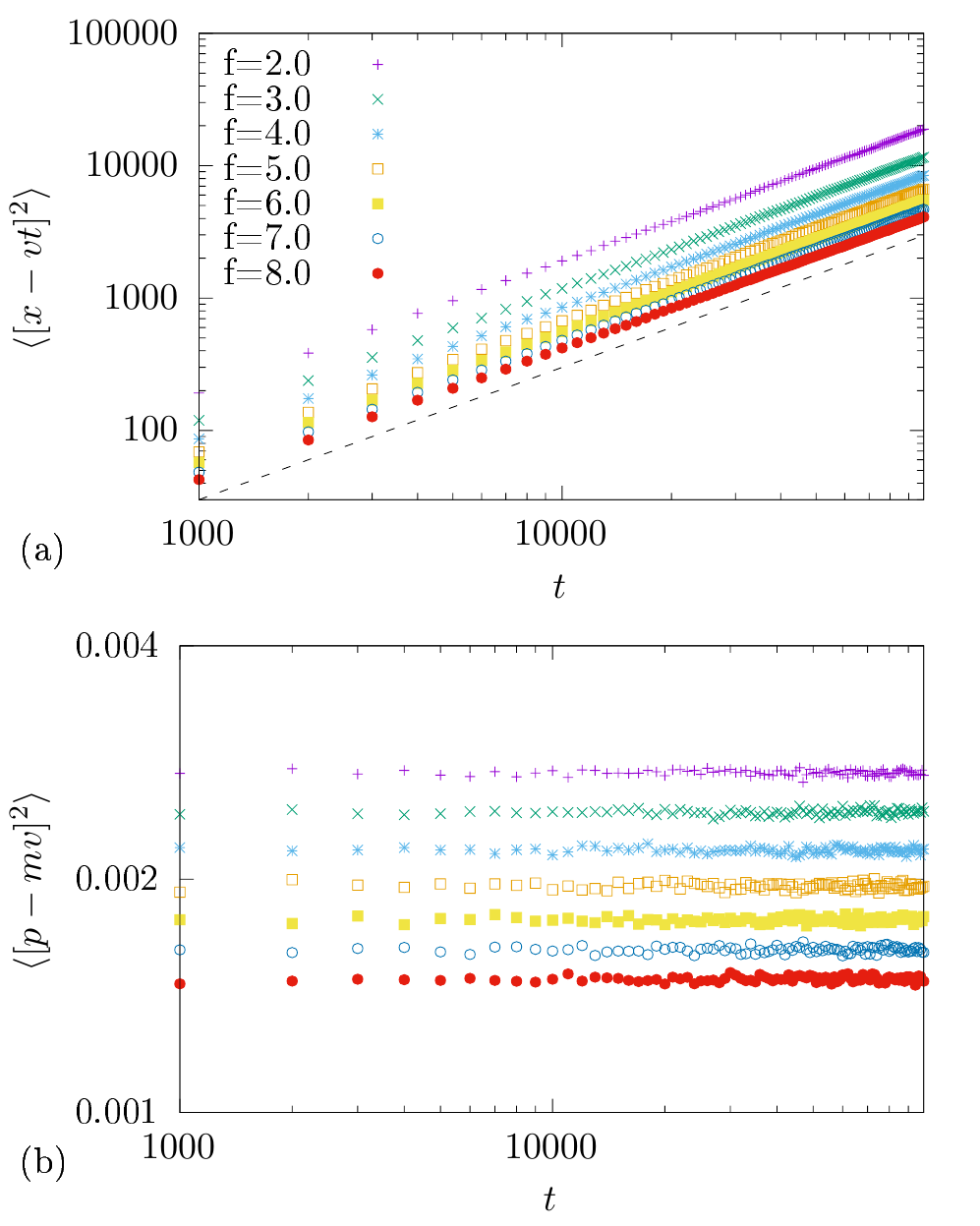}
    \caption{Quadratic mean displacement (a) and momentum variance (b) in the moving frame vs $t$ for the RF disorder case of Eq.(\ref{eq:equationwithinertia}), using $m=0.1$, and indicated driving forces. The dashed line indicates normal {dispersion}.
    }
    \label{fig:msdRF}
\end{figure}

\begin{figure}[htp]
    \centering
    \includegraphics[width=\columnwidth]{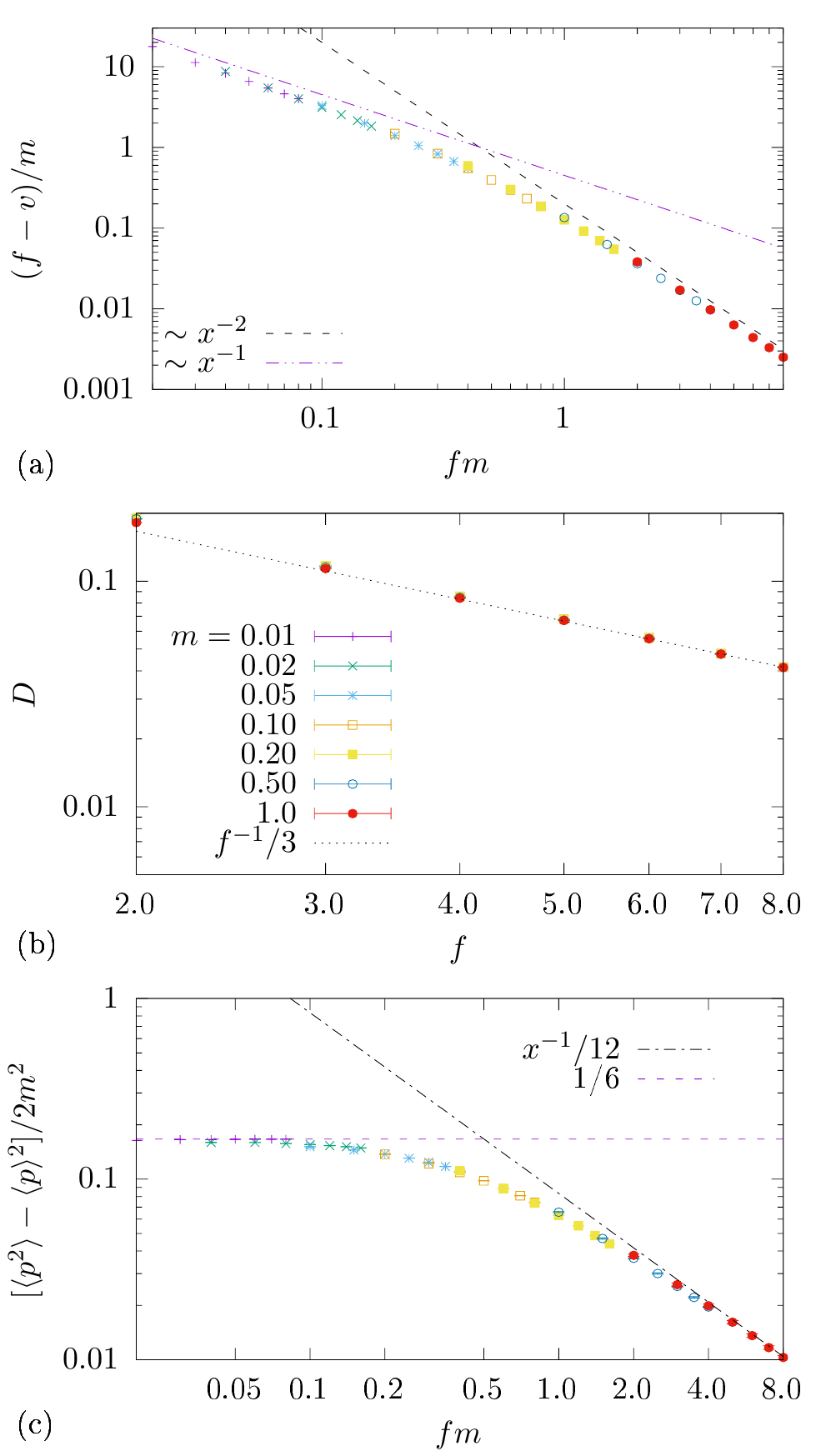}
    \caption{Numerical results for the inertial RF disorder case of Eq.(\ref{eq:equationwithinertia}). (a) Rescaled velocity deficit as a function of the drive $f$, for different masses $m$ (indicated in (b)), in the fast flow regime $f-v \ll v$. Dash and dash-dotted lines indicate approximate asymptotic behaviours. 
    (b) Dispersion constant vs $f$ for different masses. The dotted line shows a $\sim 1/f$ decay.
    (c) Rescaled momentum variance vs $f$. The dash-dotted and dashed lines show asymptotic predictions (see text).
    }
    \label{fig:figinertiaRF}
\end{figure}

We start describing the results for the RF field.
In  Fig.\ref{fig:msdRF}(a) we show the quadratic mean displacement in the moving frame $\langle [x-v t]^2\rangle$ vs $t$, where $v=\langle \dot x \rangle$ is the steady-state mean velocity. We see that dispersion is normal and thus a driving force dependent dispersion constant $D$ can be fitted. In  Fig.\ref{fig:msdRF}(b) we see that the variance of the momentum in the moving frame $\langle p-m v \rangle$, shown for the same mass and driving forces, has a well defined drive dependent steady-state value (temporal fluctuations are controlled by the number of tracks considered in the average).
In  Fig.\ref{fig:figinertiaRF}(a) we show that in the fast-flow regime 
$f-v \approx m G_v(fm)$, with $G_v(x)$ displaying approximately a crossover between two power laws, $G_v(x)\sim x^{-1}$ for $x\ll 1$ and $G_v(x)\sim x^{-3}$ for $x\gg 1$. For a fixed $f$ and a small $m$ the result is consistent with the expansion obtained for the over-damped toy model, Eq.(\ref{eq:largevexpansion}). Interestingly, there is a crossover to a different regime for a fixed value of $f m$, so the smaller $m$ the larger the crossover force $f_m\sim m^{-1}$. 
Therefore, the expansion of Eq.(\ref{eq:largevexpansion}) is not robust.  
In Fig.\ref{fig:figinertiaRF}(b) we show nevertheless that the dispersion constant $D$ is practically independent of $m$ and decays as $D \approx f^{-1}/3$, with the same power-law predicted analytically with the toy model in the RF case, Eq.(\ref{eq:RFresult}). 
In Fig.\ref{fig:figinertiaRF}(c) 
we show that the steady-state variance of the momentum obeys also a scaling $\langle p^2 \rangle-\langle p \rangle^2 \sim m^2 G_p(f m)$ with a crossover, at $f m \sim 1$, from a constant to a power law decay, $G_p(x)\approx x^{-1}/12$. Interestingly, the exact fits to $D$ and $\langle p^2 \rangle-\langle p \rangle^2$ for large $f m$ are consistent with the relation
\begin{equation}
    \frac{\langle [p-\langle p \rangle]^2 \rangle}{2m} 
    \approx \frac{D}{4} \approx \frac{T_{\tt eff}(f)}{2} 
    \label{eq:equipartitionRF}
\end{equation}
where in the last equality we have used the definition of the effective temperature from the generalized Einstein relation of  Eq.(\ref{eq:Teffdefinition}), with $\mu\approx 1$ for $f-v \ll f$.
In other words, the mean kinetic energy calculated in the moving frame with velocity $v=\langle p/m \rangle$ appears to satisfy the equilibrium equipartition law in spite of being a system far from equilibrium.
We can understand this result at the lowest order in a high velocity expansion. If we write that the random force is $F(x)\approx F(vt)$, and that $\dot x \approx v \approx f$, from Eq.(\ref{eq:equationwithinertia}) we get 
\begin{equation}
m \ddot u  \approx {F}(vt)-\dot u,
\label{eq:langevinRF}
\end{equation}
for the position coordinate $u=x-vt$ in the moving frame.
Then, using that $\overline{F(x)F(x')}=g(|x-x'|)$ with $g(x)$ a rapidly decaying function with $\int_x g(x)>0$ for RF, 
we can compute the time autocorrelation function of the random force for $v \gg |t-t'|^{-1}$,  $\overline{F(vt)F(vt')}=g(v|t-t'|)\sim \delta(t-t')/v$. 
Hence, in such large velocity limit, the random force behaves effectively as a Langevin noise $\xi_v(t)\equiv F(vt)$ \footnote{The ensemble average of the Langevin noise coincides in this limit with the average over tracks.} at an effective temperature $T_{\tt eff}\sim v^{-1}$ (since the mobility is constant and $\mu=1$). The system thus reaches thermal equilibrium at the effective-temperature $T_{\tt eff}$, and hence the mean kinetic energy is $\langle \frac{m}{2}\dot u^2 \rangle = T_{\tt eff}/2$, 
in agreement with the asymptotic result of Eq.(\ref{eq:equipartitionRF}).
On the other hand, for $fm<1$ the particle enters a regime where the friction force $\dot x$ dominates over the inertial term $m\ddot x$ 
and then, using the same arguments as above we get 
\begin{equation}
    \dot u \approx F(vt)
    \label{eq:smallfm}
\end{equation}
which implies that $\langle \dot u^2 \rangle = \langle F(vt)^2 \rangle=1/3$. Therefore,  
$\langle [p-mv]^2 \rangle/2 m^2 = 1/6$ in agreement with the scaled data shown in Fig.\ref{fig:figinertiaRF}(c) for $fm\ll 1$.

\subsubsection{RB case}
\begin{figure}[htp]
    \centering
     \includegraphics[width=\columnwidth]{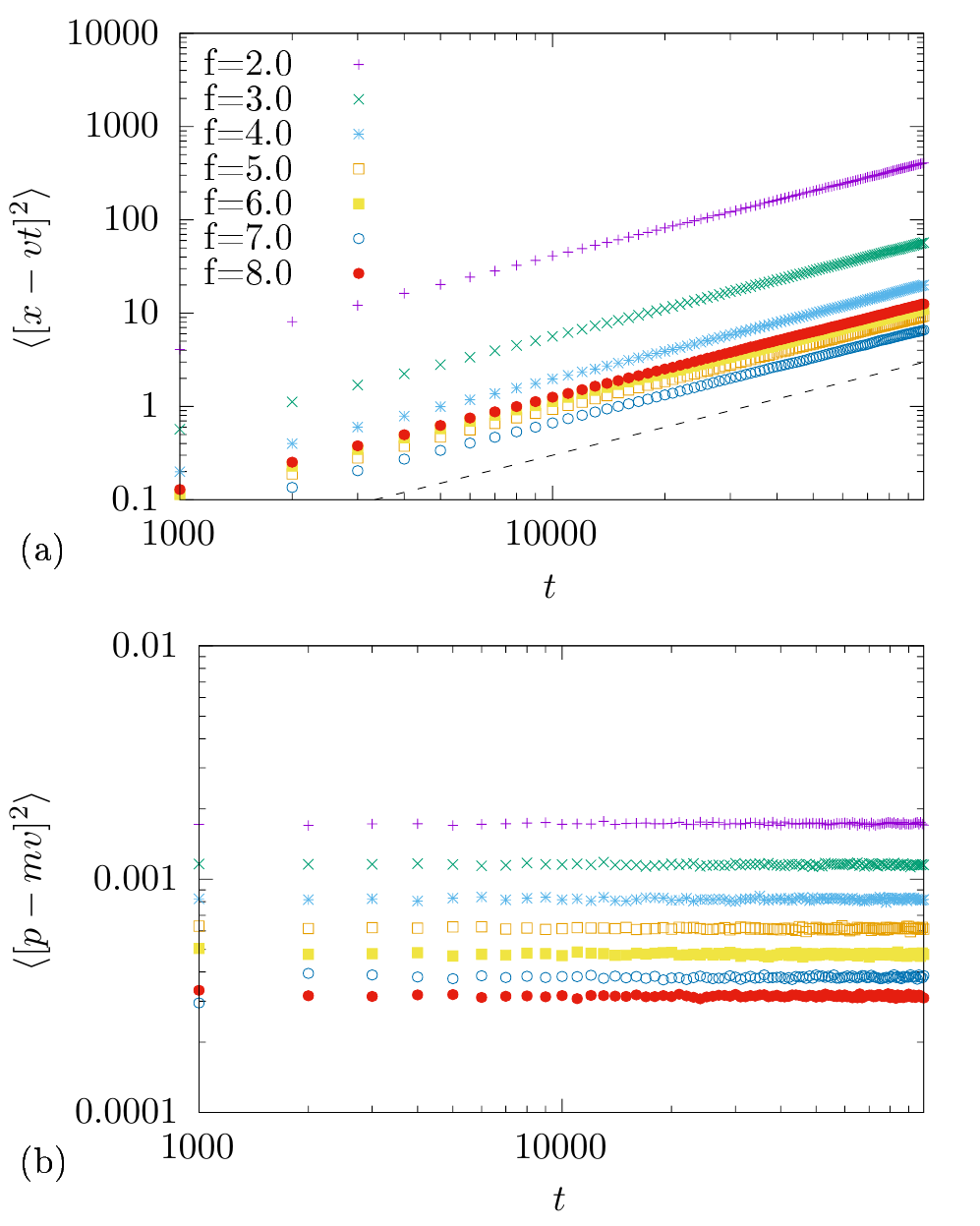}
    \caption{Quadratic mean displacement (a) and momentum variance (b) in the moving frame vs $t$, using $m=0.1$ and indicated driving forces, for the RB disorder case of Eq.(\ref{eq:equationwithinertia}). The dashed line in (a) indicates normal {dispersion}.
    }
    \label{fig:msdRB}
\end{figure}

\begin{figure}[htp]
    \centering
    \includegraphics[width=\columnwidth]{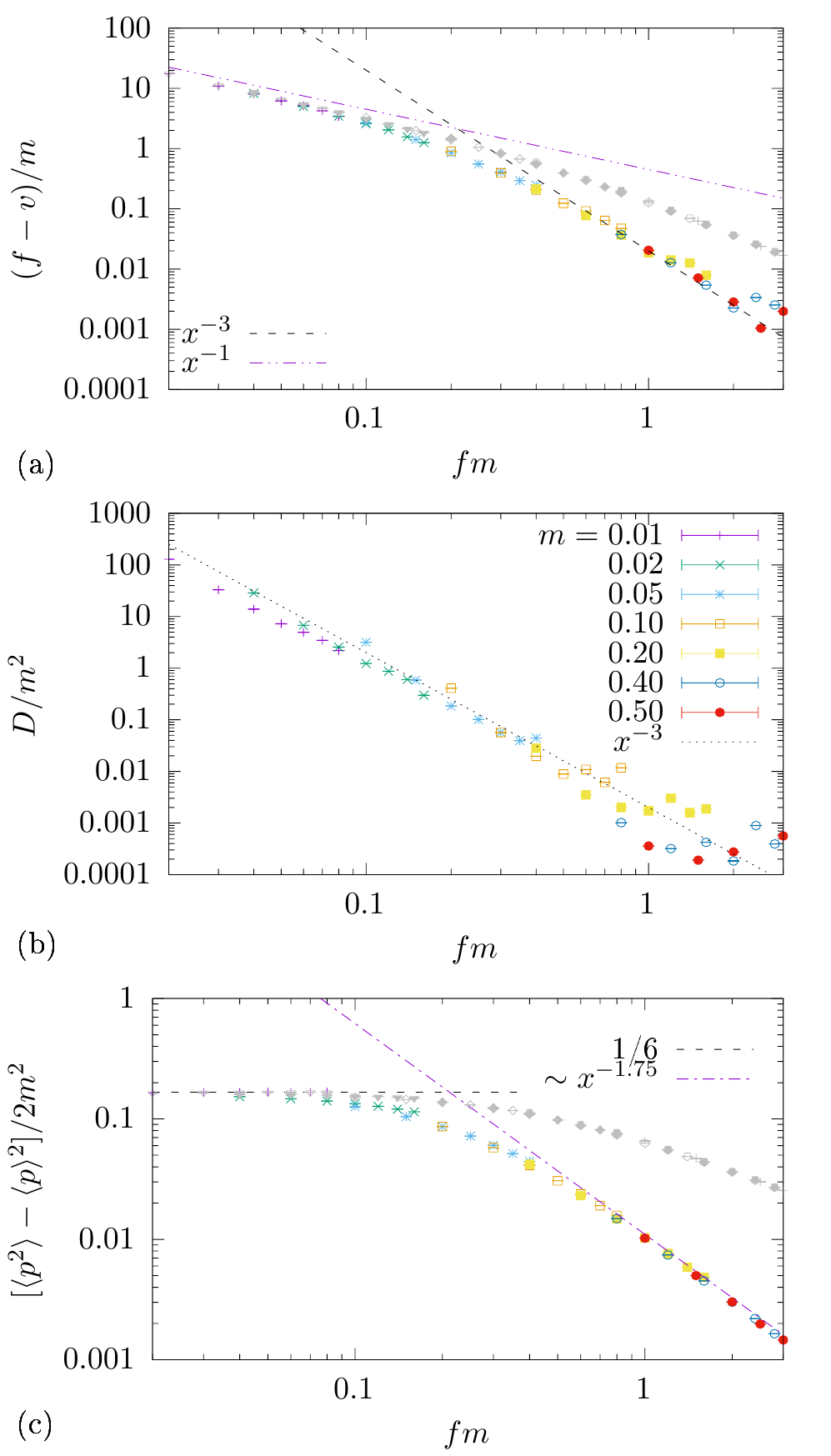}
    \caption{Numerical results for the inertial RB case of Eq.(\ref{eq:equationwithinertia}). (a) Rescaled velocity deficit as a function of the drive $f$, for different masses $m$ (indicated in (b)), in the fast flow regime $f-v \ll v$. Dash and dash-dotted lines indicate approximate asymptotic behaviours, as indicated. 
    (b) Dispersion constant vs $f$ for different masses. The dotted line highlights the $\sim f^{-3}$ decay.
    (c) Rescaled momentum variance vs $f$. The dash-dotted and dashed lines indicate approximate asymptotic behaviours. Grey symbols in (a) and (c) correspond to the results of the RF case.
    }
    \label{fig:figinertiaRB}
\end{figure}

We now describe the results for the RB field.
In  Fig.\ref{fig:msdRB}(a) we show the quadratic mean displacement in the moving frame $\langle [x-v t]^2\rangle$ vs $t$, where $v=\langle \dot x \rangle$ is the steady-state mean velocity. We see that {dispersion} is normal and thus a driving force dependent dispersion constant $D$ can be fitted. In  Fig.\ref{fig:msdRB}(b) we see that the variance of the momentum in the moving frame $\langle (p-m v)^2 \rangle$, shown for the same mass and driving forces, has a well defined drive-dependent steady-state value.
In Fig.\ref{fig:figinertiaRB}(a) we show, for various masses (indicated in Fig.\ref{fig:figinertiaRB}(b)), that the deficit of velocity with respect to the free velocity obeys  approximately the same scaling $(f-v)\sim m G_f(f m)$ found for the RF case, and also presents a crossover between two regimes. The first regime displays the correction $G_f(x)\sim x^{-1}$ expected from the large velocity expansion of the overdamped toy-model Eq.(\ref{eq:largevexpansion}). The second regime $G_f(x)\sim x^{-2}$ for $f m \gg 1$, shows that inertia can have an important effect, and thus the scaling of Eq.(\ref{eq:largevexpansion}) is not universal. 
In Fig.\ref{fig:figinertiaRB}(b) we show nevertheless that the dispersion coefficient $D$ follows closely the $f^{-3}$ predicted by the toy model in the RB case, Eq.(\ref{eq:RBresult}).
It is nevertheless worth noting that while in the RF case $D$ is almost independent of $m$, in the RB case there is a strong dependence, as evidenced by the rescaling of curves into a master curve.
We also note that $D$, for the same range of $m$ and $f$, is in general much smaller in the RB case than in the RF case, and more difficult to sample precisely.
Finally, in Fig.\ref{fig:figinertiaRB}(c) 
we show that the momentum variance follows the same scaling $\langle p^2\rangle-\langle p\rangle^2 = m^2 G_p(fm)$ than in the RF case, and a similar crossover from small to large $fm$. Interestingly, the small $fm$ regime saturates to an identical value than for the RF case. This is consistent with the effective equation of motion Eq.(\ref{eq:smallfm}), with the saturation value controlled only by the variance of the disorder. 
The large $fm$ regime however, where empirically we find $G_p(x)\sim x^{-1.75}$, differs appreciably from the RF case. The faster decay of 
$G_p(x)$ is however consistent with the faster decay of $G_f(x)$ 
shown in Fig.\ref{fig:figinertiaRB}(a). 
At variance with the RF case, The large $fm$ regime can not be rationalized using an effective equipartition theorem  because $D \sim T_{\tt eff}$ does not decay in the same way in this limit. 
The interpretation of this remains open. 

\subsubsection{Summary}
In summary, we find that damped particles display:
\begin{itemize}
    \item $D\sim 1/f$ and $D\sim 1/f^3$ as predicted for the overdamped toy model for the RF and RB case respectively, Eqs.(\ref{eq:RFresult}) and (\ref{eq:RBresult}). The result thus appears to be robust under inertia.
    \item A non trivial dependence with the mass $m$ both in the $f-v$ and in $\langle p^2 \rangle-\langle p \rangle^2$ reveals a crossover between two regimes varying the parameter $f m$. The first regime, identical for RB and RF is well described by the overdamped limit with an effective drive-dependent noise. The second regime is different for RF and RB. In the RF case the second regime is compatible with a generalized equipartition law at the disorder-induced drive dependent effective temperature $T_{\tt eff}=D/2$. For RB the later equipartition law {\it fails}. 
\end{itemize}

\subsection{Quasi one- and two-dimensional tracks}
\label{sec:transverse}
When the size of the particles is smaller than the track width transverse fluctuations of a transversely confined particle may modify the predicted longitudinal dispersion properties. This kind of situation may arise for instance in the cases of driven vortices in narrow and thin superconducting strips, driven colloids,  or current driven magnetic skyrmions. Compared to the simple one-dimensional overdamped case, now the state-space doubles (as in the one-dimensional damped case) and the effective one-dimensional coupling to the underlying disorder changes.
We study the consequences of these new properties in the velocity and dispersion of particles. 

\subsubsection{RB case}
\begin{figure}[htp]
    \centering
    \includegraphics[width=\columnwidth]{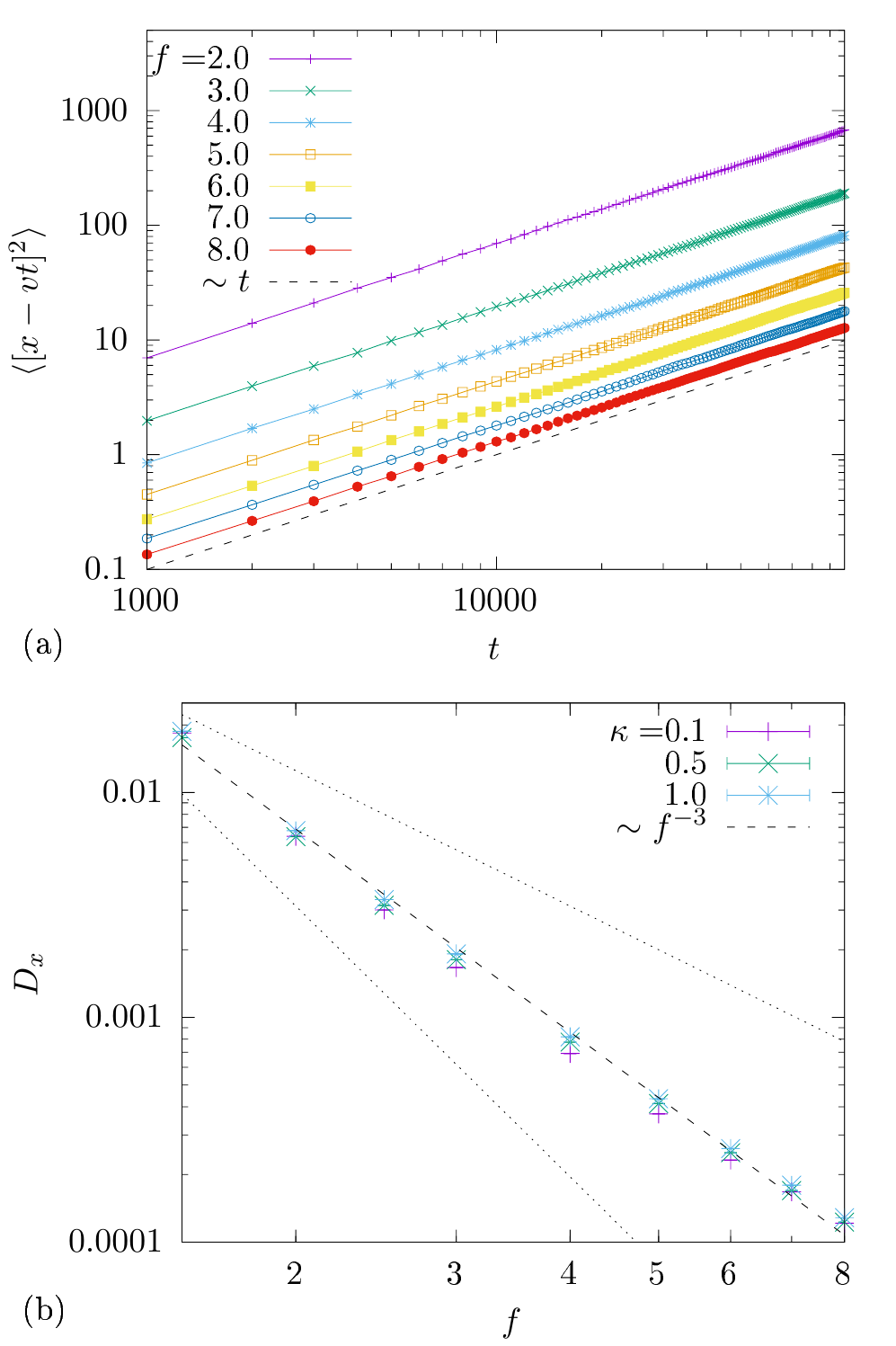}
    \caption{
    RB results for the two dimensional model of Eq.(\ref{eq:2d}) with transverse confinement in a  two-dimensional random bounded potential.
    (a) Quadratic mean displacement with respect to the mean motion in the driving direction vs time, for different driving forces $f$ for a confinement constant $\kappa=1$. The dashed line indicates normal dispersion.
    (b) Fitted dispersion constant in the direction of the drive $D_x$ vs $f$ for different confinement constants $\kappa$. The dashed line indicates the $1/f^3$ behaviour. For comparison $1/f^2$ and $1/f^4$ behaviors are drawn in dotted lines.
    }
    \label{fig:figconfined1}
\end{figure}

\begin{figure}[htp]
    \centering
    \includegraphics[width=\columnwidth]{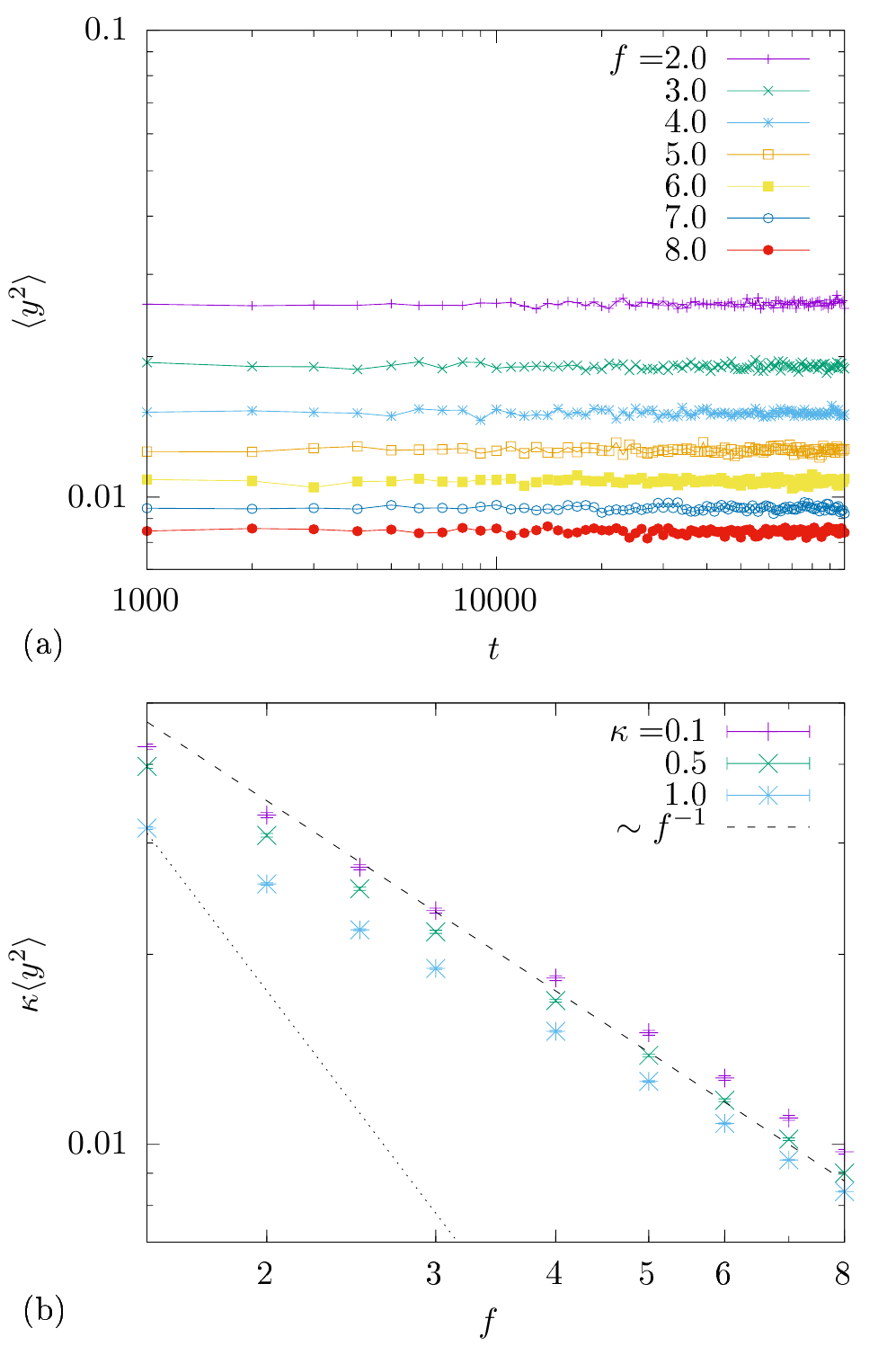}
    \caption{
    RB results for the two dimensional model of Eq.(\ref{eq:2d}) with transverse confinement ($\kappa>0$) for the  two-dimensional uncorrelated random potential of Eq.(\ref{eq:bilinear}). 
    (a) Mean squared displacement in the transverse direction as a function of time, for $\kappa=1$.
    (b) The variance of transverse fluctuations approximately satisfies $\kappa \langle y^2 \rangle \sim f^{-1}$  at large forces for different confinement constants $\kappa$. Dotted line indicate the $1/f^2$ behavior for comparison.}
    \label{fig:figconfined2}
\end{figure}

\begin{figure}[htp]
    \centering
    \includegraphics[width=\columnwidth]{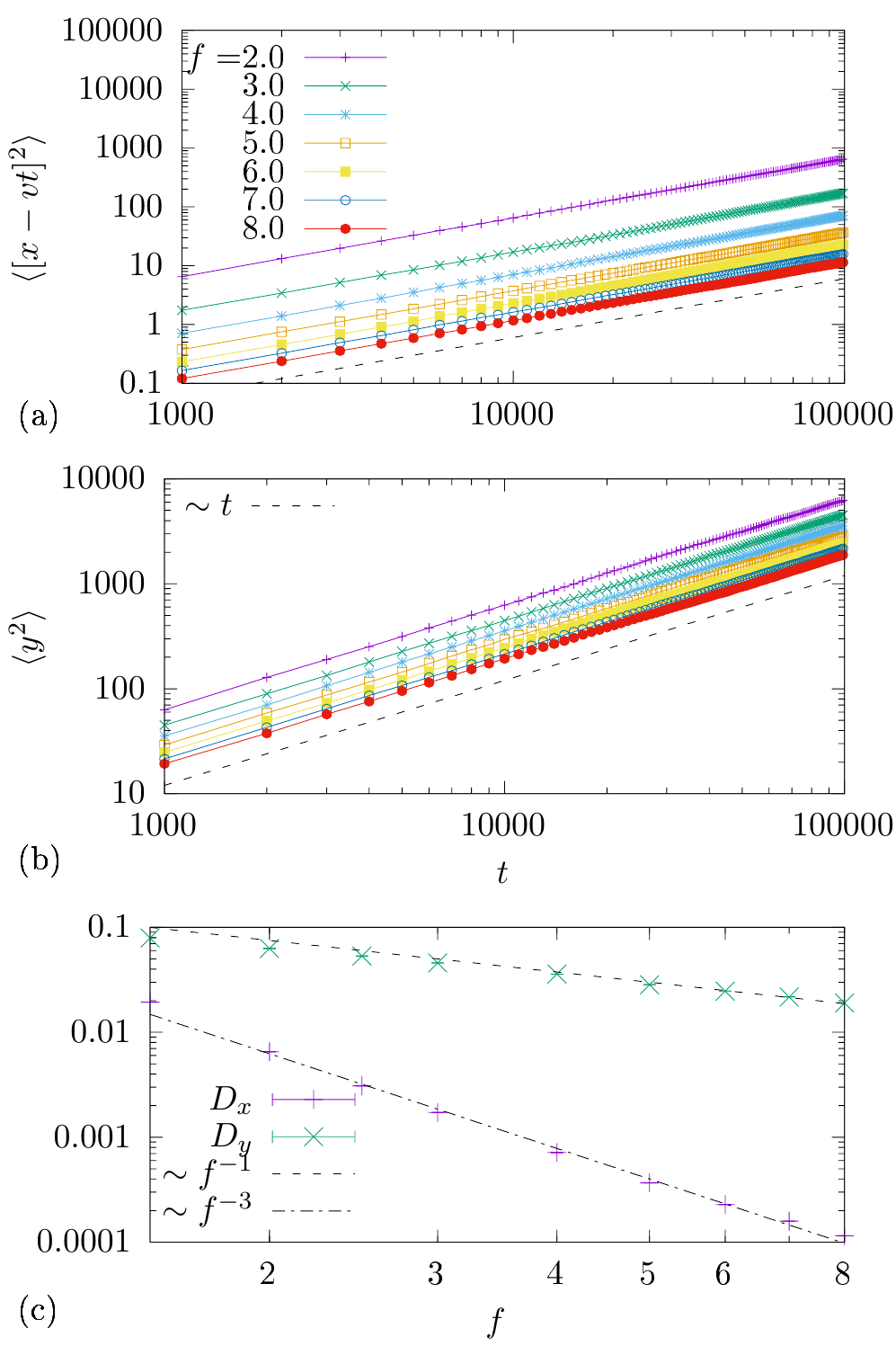}
    \caption{
    RB results for the two dimensional model (Eq.(\ref{eq:2d})) \textit{without} confinement ($\kappa=0$), for the  two-dimensional uncorrelated random potential of Eq.(\ref{eq:bilinear}).
    Longitudinal (a) and transverse (b) quadratic mean displacement vs time for different driving forces $f$, with respect to the mean motion.
    (c) Longitudinal and transverse dispersion constants $D_x$ and $D_y$ as a function of $f$. The dashed line shows the $\sim f^{-1}$ and the dash-dotted the $\sim f^{-3}$ behaviours.
    }
    \label{fig:figconfined3}
\end{figure}

In order to study the effect of transverse fluctuations we will consider here only the relevant RB case of an overdamped particle in a narrow channel transversely confined by a parabolic potential 
\begin{eqnarray}
\dot x &=& f - \partial_x U(x,y), 
\nonumber \\
\dot y &=& -\kappa y -\partial_y U(x,y),
\label{eq:2d}
\end{eqnarray}
with $\kappa \geq 0$ and $U(x,y)$ a random potential in the plane, obtained by bilinear interpolation, 
such that for $x \in [n,n+1]$ and $y \in [m,m+1]$ 
we have
\begin{eqnarray}
U(x,y)&=& U_{n,m}(n+1-x)(m+1-y) \nonumber\\
&+& U_{n+1,m}(x-n)(m+1-y) \nonumber\\
&+& U_{n,m+1}(y-m)(n+1-x) \nonumber\\ 
&+& U_{n+1,m+1}(y-m)(x-n),
\label{eq:bilinear}
\end{eqnarray}
where $U_{n,m}$ are quenched independent random values, $\langle U_{n,m} \rangle=0$ and $\langle U_{n,m}U_{n',m'} \rangle=\delta_{n,n'}\delta_{m,m'}/3$,  drawn from a uniform distribution.
In Fig.\ref{fig:figconfined1} we show results for $\kappa=1$. The quadratic mean displacement with respect to the mean motion at velocity $v$ in the direction of the drive (in the other direction the mean velocity vanished by symmetry) grows linearly with time, as indicated in Fig.\ref{fig:figconfined1}(a), and dispersion is normal with different $f$-dependent dispersion constants $D_x$, which are fitted as $\langle [x-vt]^2 \rangle \sim D_x t$ at long times. In Fig.\ref{fig:figconfined1}(b) we show that $D_x \sim 1/f^3$ at large $f$, being almost independent of the confinement in one decade of variation of the parameter $\kappa$. This dependence with $f$ is consistent with the one predicted for the one-dimensional system in a RB force field. In Fig.\ref{fig:figconfined2}(a) we show that the mean quadratic transverse displacement $\langle y^2 \rangle$ vs time $t$ saturates at a well defined $f$-dependent value. Interestingly, we find that $\kappa \langle y^2 \rangle \sim f^{-1}$, as indicated in Fig.\ref{fig:figconfined2}(b).

The result of Fig.\ref{fig:figconfined2}(b) can be interpreted as an equipartition law analog in the transverse direction, with an effective transverse temperature vanishing as $T^y_{\tt eff}\sim f^{-1}$, at variance with the longitudinal fluctuations that vanish as $D_x \sim T^x_{\tt eff} \sim f^{-3}$. 
To further test this picture in Fig.\ref{fig:figconfined3} we show the unconfined $\kappa=0$ case, for the longitudinal (a) and transverse (b) fluctuations. As can be observed, in this case transverse fluctuation lead to normal transverse dispersion. The dependencies $D_x\sim 1/f^3$ and $D_y\sim 1/f$ are consistent with the picture of two different temperatures, one for each direction. These temperatures satisfy a generalized fluctuation-dissipation relation analogous to the Einstein relation between the temperature and the mobility of a Brownian particle, since the differential mobility in each direction becomes drive independent at large $v$.  This kind of behaviour was already observed in a two-dimensional toy models with randomly located parabolic wells~\cite{Kolton2006}. A simple (but not general) argument starts by noting that in the particular bilinearly interpolated potential transverse force in a given square cell derived from Eq.(\ref{eq:bilinear}) is independent of $y$, i.e. $-\partial_y U(x,y)\equiv F_y(x)$. In the next cell the same holds but the force $F_y(x)$ changes, because the random potential at the vertex of the unit squares $V_{n,m}$ are uncorrelated quenched random numbers. 
Since $\langle \partial_y U(x,y)\partial_y U(x',y')\rangle=\Delta_{yy}(x-x',y-y')$,
at large $v$ we can write $\langle F_y(x(t))F_y(x(t')) \rangle \approx \Delta_{yy}(x(t)-x(t'),y=0) \approx \Delta_{yy}(v(t-t'),y=0) \sim \delta(t-t')/v \equiv 2T^y_{\tt eff}(v)\delta(t-t')$, where in the first approximate equality we used that the particle moves mostly forward in a distance equal to one correlation length of the disorder along $x$, and in the fourth term that $\Delta_{yy}$ is short-ranged and satisfies $\int_u \Delta_{yy}(u) \neq 0$.
Therefore, we see that the transverse force mimics a thermal-noise at an effective transverse temperature vanishing as $T^y_{\tt eff} \sim 1/f \sim 1/v$. 
A more general argument to understand the $1/f$ transverse fluctuations in the RB case at large velocities is given in Ref.\cite{Elias2022}, in the context of a vortex line dynamics.

\subsubsection{RF case}
\begin{figure}[htp]
    \centering
    \includegraphics[width=\columnwidth]{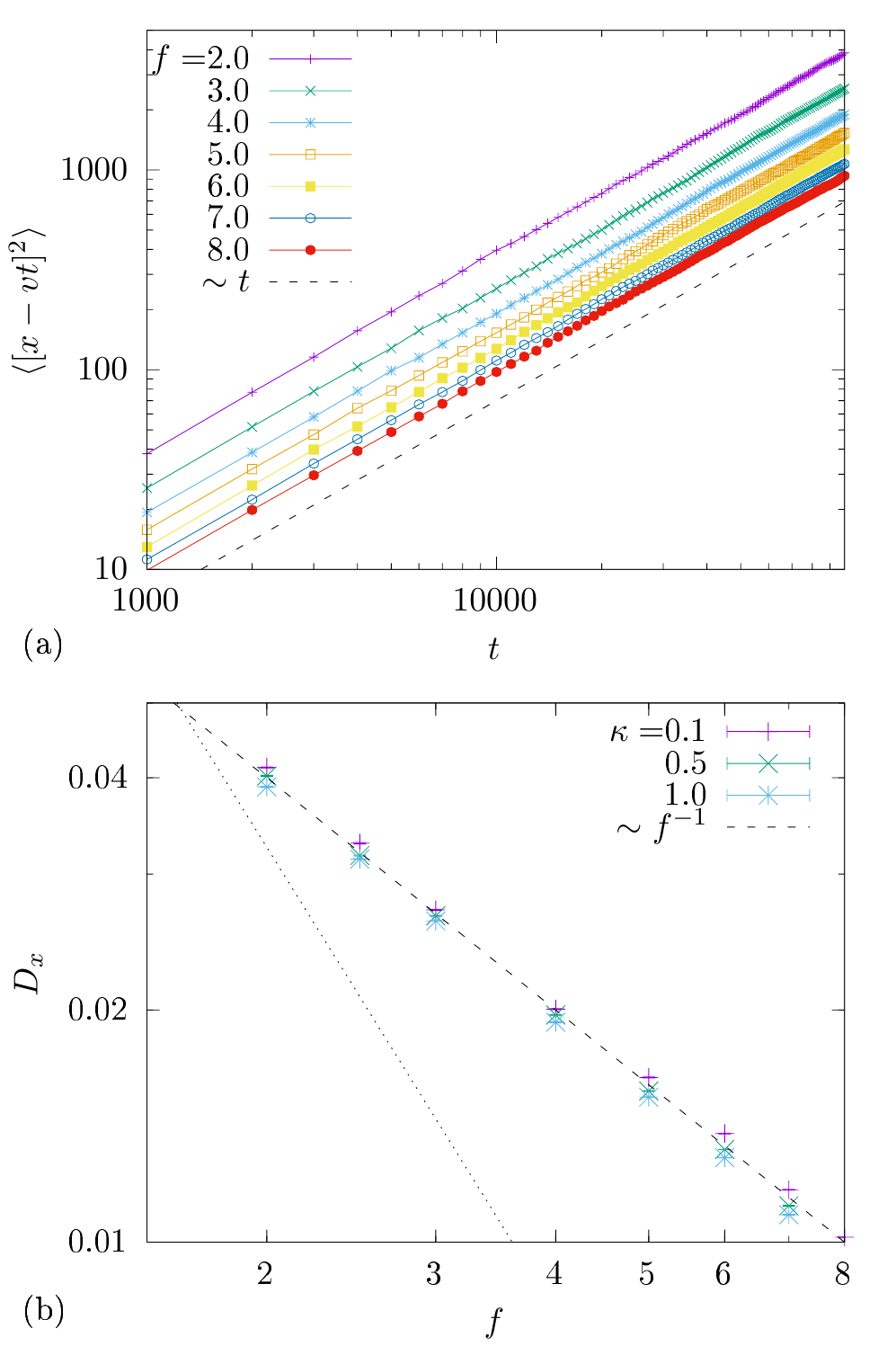}
    \caption{
    RF results for the two dimensional model  with transverse confinement, Eq.(\ref{eq:2d}).
    (a) Quadratic mean displacement vs time with respect to the mean motion in the driving direction, for different driving forces $f$ for a confinement constant $\kappa=1$. The dashed line indicates normal dispersion.
    (b) Fitted dispersion constant in the direction of the drive $D_x$ vs $f$ for different confinement constants $\kappa$. The dashed line indicates the $1/f$ decay and dotted line indicates the $1/f^2$ behavior for comparison.
    }
    \label{fig:figconfined1RF}
\end{figure}

\begin{figure}[htp]
    \centering
    \includegraphics[width=\columnwidth]{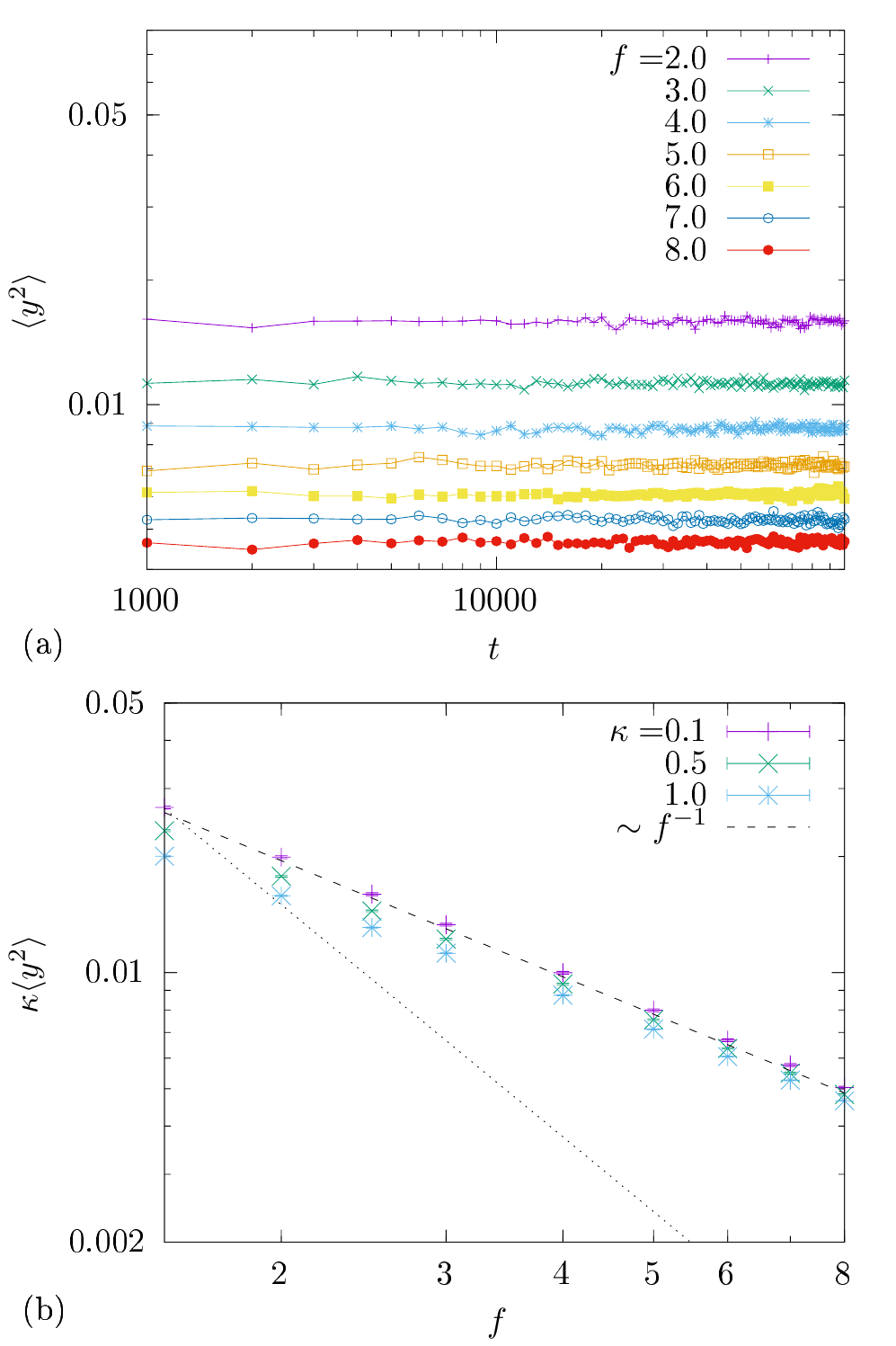}
    \caption{
    RF results for the two dimensional model  with transverse confinement $\kappa>0$, Eq.(\ref{eq:2d}). 
    (a) Mean squared displacement in the transverse direction as a function of time, for $\kappa=1$.
    (b) The variance of transverse fluctuations approximately satisfies $\kappa \langle y^2 \rangle \sim f^{-1}$  at large forces for different confinement constants $\kappa$.  Dotted line indicates the $1/f^2$ behavior for comparison.
    }
    \label{fig:figconfined2RF}
\end{figure}

\begin{figure}[htp]
    \centering
    \includegraphics[width=\columnwidth]{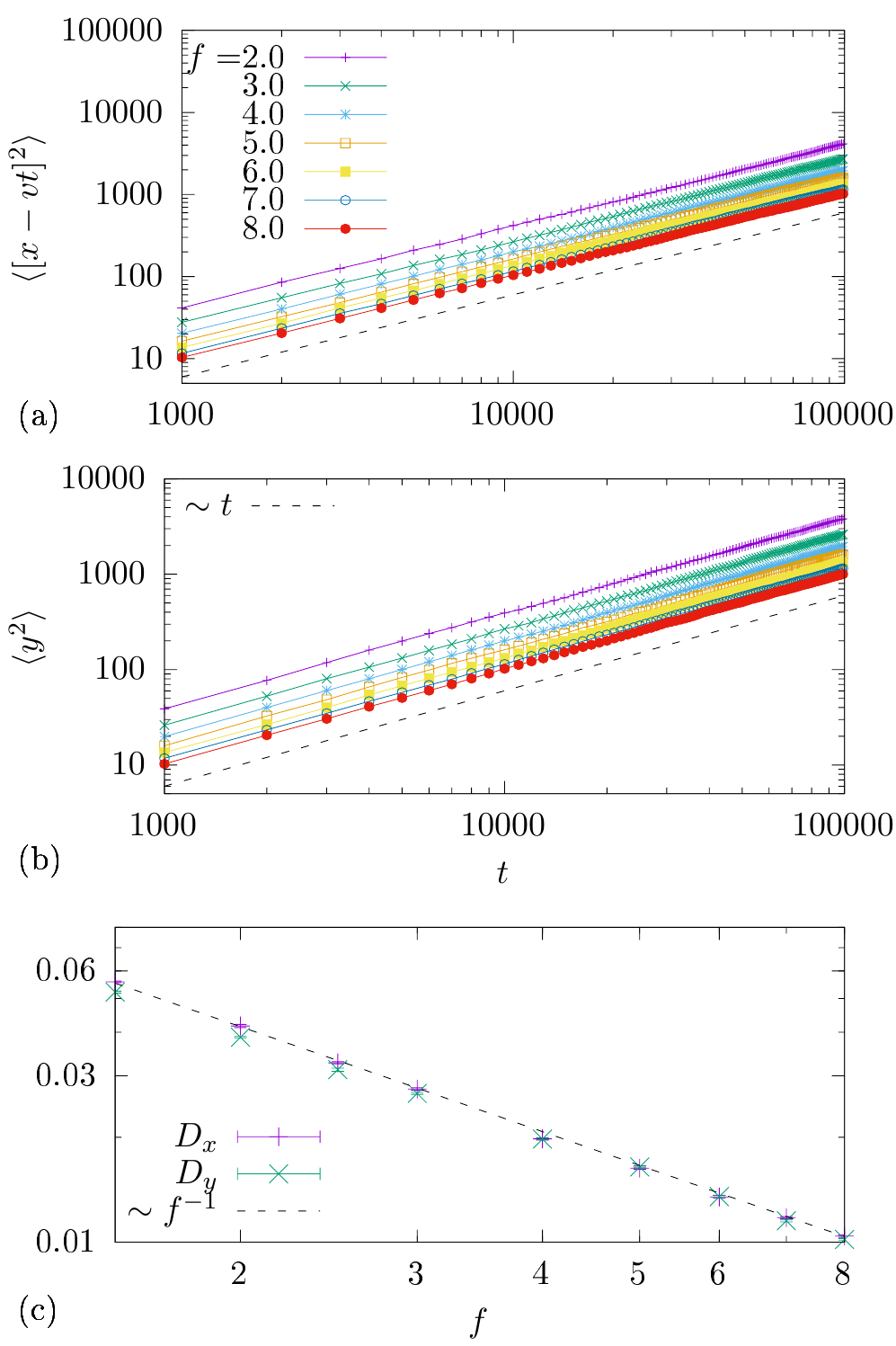}
    \caption{
    RF results for the two dimensional model  \textit{without} confinement ($\kappa=0$), from Eq.(\ref{eq:2dRF}).
    Longitudinal (a) and transverse (b) quadratic mean displacement vs time for different driving forces $f$, with respect to the mean motion. Dashed lines in (a) and (b) highlight normal dispersion.
    (c) Longitudinal and transverse dispersion constants $D_x$ and $D_y$ as a function of $f$. The dashed line shows the $\sim f^{-1}$ decay.
    }
    \label{fig:figconfined3RF}
\end{figure}

In two dimensions the RF case contemplates various possible cases. The random force field can be compressible or incompressible, rotational or irrotational or a mixture of components with different properties. Here we consider a simple one, which is to take 
$\overline{F_x(x,y)F_y(x',y')}=0$
and $\overline{F_x(x,y)F_x(x',y')}=\overline{F_y(x,y)F_y(x',y')}=\Delta(|x-x'|)\Delta(|y-y'|)$, with $\Delta(z)$ a positive short-ranged function. 
Note that, unlike the $d=1$ RF case, this force can not be derived from a potential i.e. ${\bf F}({\bf x})\neq -\nabla U({\bf x})$. 
In practice, for coordinates  
$x \in [n,n+1]$ and $y\in (m,m+1)$ 
in cell $(n,m)$ we generate two uniformly distributed independent random numbers in the $(-1/2,1/2)$ interval, one for $F_x(x,y)$ and the other for $F_y(x,y)$~\footnote{Different choices, such as $F_x=F_0 \cos(\alpha_{n,m})$, $F_y=F_0 \sin(\alpha_{n,m})$, with $\alpha_{n,m}\in (0,2\pi)$ a random angle, yield equivalent results.}. 
The equations of motion then read
\begin{eqnarray}
\dot x &=& f + F_x(x,y), 
\nonumber \\
\dot y &=& -\kappa y +F_y(x,y).
\label{eq:2dRF}
\end{eqnarray}
In Fig.\ref{fig:figconfined1RF}(a) we show the quadratic mean displacements in the moving frame vs time. Dispersion in the longitudinal direction is normal and drive-dependent. Comparing with Fig.\ref{fig:figconfined1} for RB we can see a larger dispersion for the RF case. This is reflected in a larger dispersion coefficient $D_x$ in Fig.\ref{fig:figconfined1RF}(b), where we can also observe a $D_x\sim f^{-1}$ decay at large $f$, at variance with the RB case, but consistent with the prediction of the one dimensional toy model, Eq.(\ref{eq:RFresult}). We can also observe that this result is almost independent of the confinement constant $\kappa$ in Eq.(\ref{eq:2dRF}),  as was also observed in the RB case.
In Fig.\ref{fig:figconfined2RF}(a) we show the behaviour of the transverse fluctuation vs time, for $\kappa>0$. 
As expected, it has a well defined temporal mean value, which is drive dependent. In Fig.\ref{fig:figconfined2RF}(b) we show, for three values of $\kappa$, that  $\kappa \langle y^2 \rangle \sim f^{-1}$, as for the RB case.
Interestingly, for the unconfined two dimensional case, $\kappa=0$, the dispersion is {\it isotropic} in the commoving frame, as can be appreciated in Fig.\ref{fig:figconfined3RF}(c), which is obtained by fitting the quadratic mean displacements of Figs. \ref{fig:figconfined3RF}(a) and (b).

\subsubsection{Summary}
In summary in the two-dimensional case (either confined or unconfined)
we find that:
\begin{itemize}
    \item $D\sim 1/f^3 \sim 1/v^3$ for RB and 
$D\sim 1/f \sim 1/v$ for RF,
at large $v$.
This shows that the results for the toy model, Eqs.(\ref{eq:RFresult}) and (\ref{eq:RBresult}) respectively, are robust. This is somehow expected for $\kappa \gg 1$, as the system becomes quasi one-dimensional. Nevertheless, we have also shown that it even holds for the truly two-dimensional $\kappa=0$ case.

\item In the $\kappa=0$ case there is a normal transverse dispersion with dispersion constant decaying $D_y\sim f^{-1}$, both for RF and RB. Nevertheless dispersion in the moving frame is highly anisotropic in the RB case while it is isotropic in the RF case. This striking difference can be used as a  fingerprint of the underlying disorder.

\item Transverse fluctuations in the confined case $\kappa>0$, which decay as $\kappa \langle y^2 \rangle \sim f^{-1}$, are consistent with the effective temperature $T^{x}_{\tt eff}=D_x/2$ obtained from the unconfined $\kappa=0$ case.
\end{itemize}
 
 \subsection{Soft particles: elastic dimer model}

Strictly, skyrmions, vortices, walls or solitons in a one-dimensional disordered track are not rigid objects, and hence quenched disorder affects not only the longitudinal dispersion of their centers of mass but also induce  shape fluctuations which in turn affect the way the deformable object couples to the underlying quenched disorder. 
To model this effect, and check the robustness of dispersion properties, we will consider a model of a soft localized object composed by two particles coupled with a spring  of natural size $l_0$ in a one dimensional force field $F(x)$,
\begin{eqnarray}
\dot{x_1}&=&f - \kappa (x_1-x_2+l_0) + F(x_1), \nonumber \\  \dot{x_2}&=&f + \kappa (x_1-x_2+l_0) + F(x_2),
\label{eq:coupled}
\end{eqnarray}
where $\kappa>0$ a coupling constant controlling the (longitudinal) shape  fluctuations, allowing us to go from the rigid object for $\kappa \gg 1$ to the soft limit for $\kappa \ll 1$. 
For the simulations we use the same piece-wise $F(x)$ considered in sections \ref{sec:boxdist} for the RB and RF cases. We will be interested in the dispersion of the center of mass of the two particles.
\begin{figure}[htp]
    \centering
    \includegraphics[width=\columnwidth]{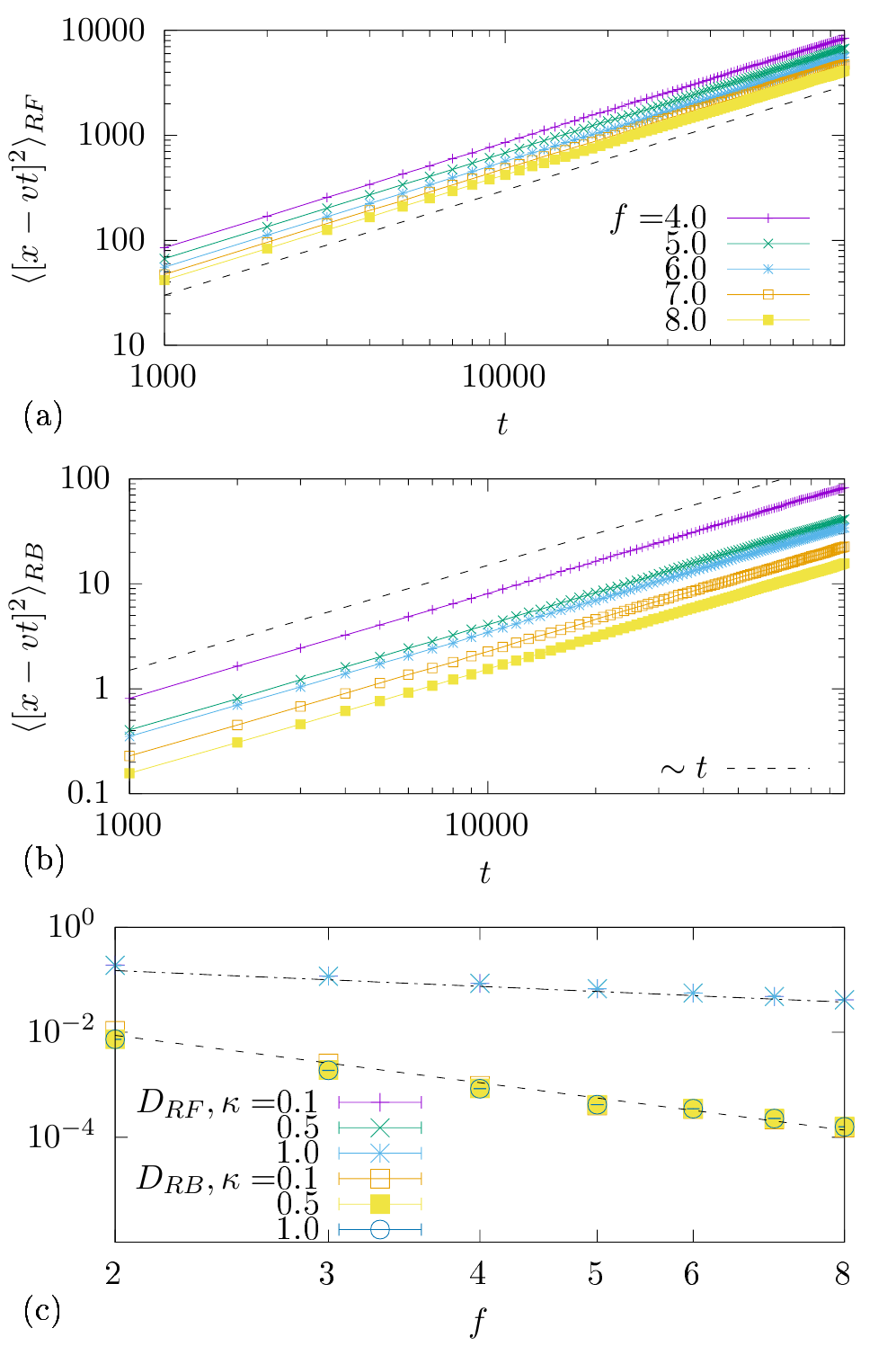}
    \caption{
    Quadratic mean displacement vs time of two coupled particles in a one dimensional track, Eq.(\ref{eq:coupled}), in the RF (a) and RB (b) disorder cases, for different driving forces. The dashed lines show that dispersion is normal in both cases. (c) The corresponding dispersion constants, for different spring constants $\kappa$, vanish as $D_{RF}\sim 1/f \sim 1/v$ (dash-dotted line)
    and $D_{RB}\sim 1/f^3 \sim 1/v^3$ (dashed line) at large $v$.
    }
    \label{fig:figcoupled}
\end{figure}

\begin{figure}[htp]
    \centering
    \includegraphics[width=\columnwidth]{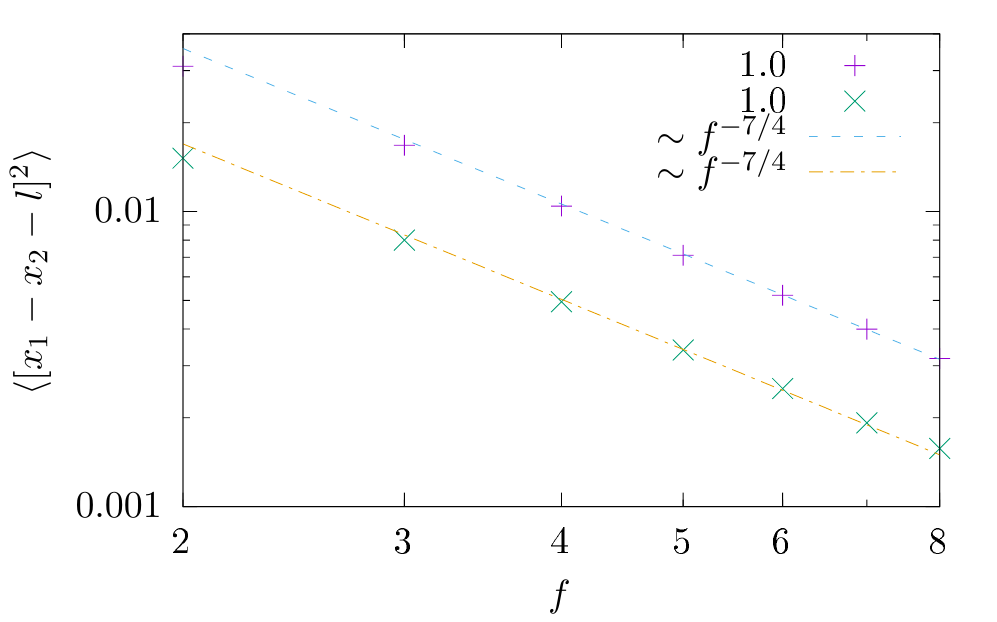}
    \caption{
    Bond fluctuations vs $f$ for two coupled particles, from Eq.(\ref{eq:coupled}), in the RF and RB cases for spring constant $\kappa=1$. As indicated by the dashed and dash-dotted lines both cases follow an empirical $\approx 1/f^{7/4}$ decay (with different amplitude), contrasting with the different decays $D_{RF}\sim 1/f$ and $D_{RB} \sim 1/f^3$ (RB) observed in the dispersion constant of individual particles (see Fig.\ref{fig:figcoupled}(c)).
    }
    \label{fig:figbond}
\end{figure}
In Fig.\ref{fig:figcoupled} (a)-(b) we show the quadratic mean displacement in the RF and RB pinning fields respectively. In both cases dispersion is normal and the dispersion constant decays with increasing $f$. In Fig.\ref{fig:figcoupled} (c) we see that $D_{RF}\sim 1/f$ and $D_{RB}\sim 1/f^3$, as obtained for the toy particle model (Eq.(\ref{eq:RFresult}) and (\ref{eq:RBresult}) respectively), for various values of the spring constant $\kappa$. This shows again the robustness of these predictions. 
Interestingly, as shown in Fig.\ref{fig:figbond}, we find that the bond fluctuations, measuring the mean Hook energy of the pair, follows the  same $\sim 1/f^{7/4}$ decay at large $f$, though with a different prefactor. This shows on one hand that the effective equipartition law that approximately holds for the transverse fluctuations of transversely confined particles (see section \ref{sec:transverse}) does not work in this case, in the sense that bond fluctuations are clearly not controlled by the same effective temperature $T_{\tt eff}\approx D/\mu$ describing the center of mass dispersion.
This may be attributed to the fact that the effective disorder-induced noise is correlated between particles, because they follow each other and thus visit exactly the same quenched disorder after a mean characteristic time $\sim l_0/v$. Hence, this noise can not mimic thermal-like noise $\xi_n(t)$, which must satisfy $\langle \xi_n(t)\xi_m(t') \rangle \propto \delta_{n,m}\delta(t-t')$ for particles $n$ and $m$ in a coarse-grained level. It is nevertheless quite remarkable that the RB and RF cases seem to follow a unique decay law for bond fluctuations, in sharp contrast with $D_{RF}$ and $D_{RB}$ for the center of mass. It would be interesting to investigate if this system can be still be described by multiple effective temperatures, depending on the typical time-scale of the observable fluctuations.

\subsubsection{Summary}
In summary, we have shown that a simple model of a simple deformable object (an elastic dimer) driven in a disordered one-dimensional track, presents at large velocity a dispersion described by Eqs.(\ref{eq:RFresult}) and (\ref{eq:RBresult}) made for the one overdamped toy model. These predictions thus appear robust under shape fluctuations of soft but yet localized objects.

\subsection{Effect of temperature: driven Brownian particle}
\label{sec:langevin}
\begin{figure}[htp]
    \centering
    \includegraphics[width=\columnwidth]{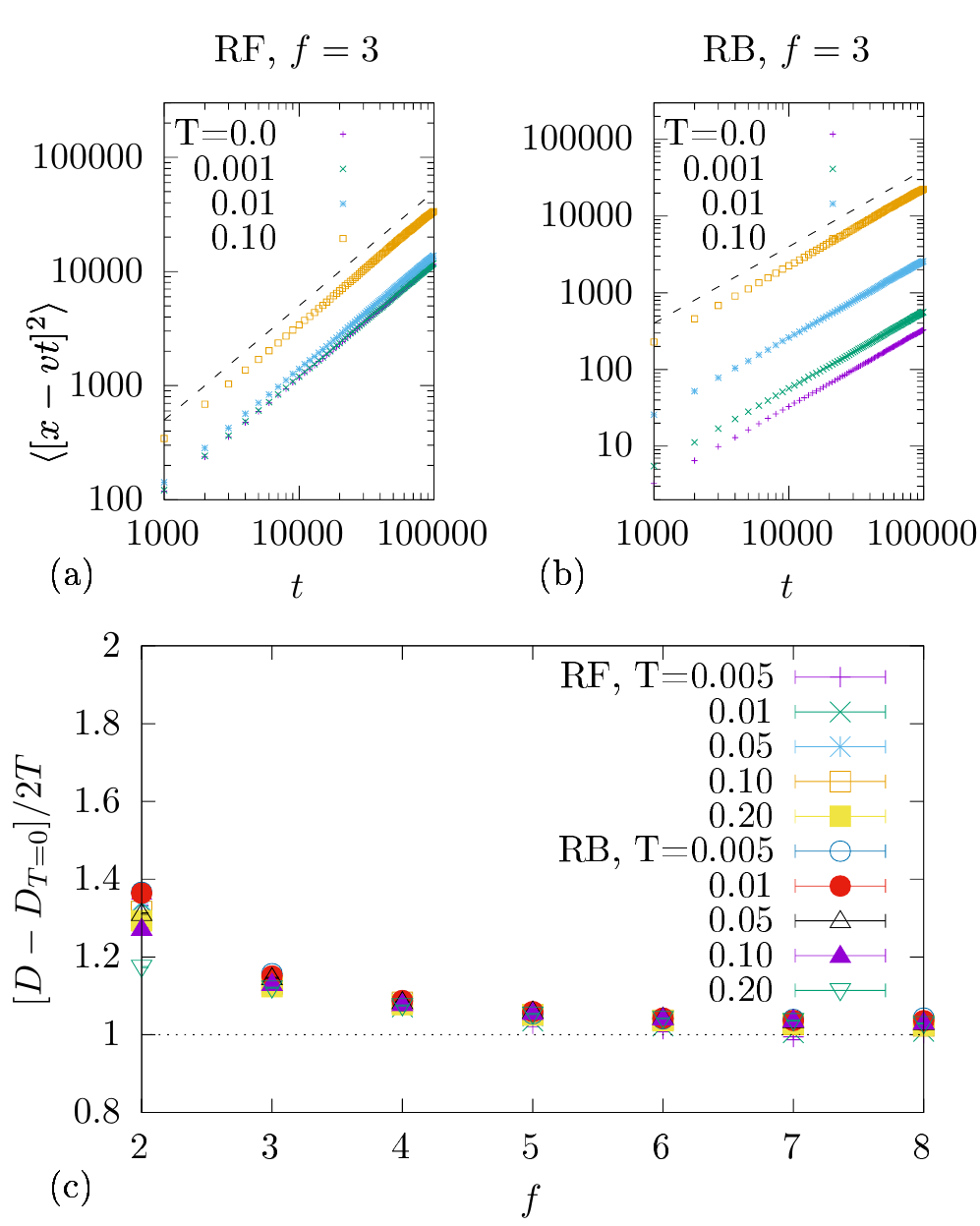}
    \caption{
    Effect of temperature on the dispersion in the one dimensional overdamped particle model.
    Quadratic mean displacement vs time for different temperatures in the RF (a) and RB (b) random force fields, for a given driving force $f=3$ (dashed lines indicate normal dispersion).
    (c) At large $f$, thermal noise becomes just an additive effect in the dispersion, and $D(f,T)\approx D(f,T=0)+2T$ (as indicated by the asymptotic dotted line) independently of the RF or RB pinning nature.  
    }
    \label{fig:temperature}
\end{figure}

So far we have discussed deterministic dispersion of different objects in narrow tracks which can be described as particles with one or more degrees of freedom.
When the dispersing objects are small enough thermal fluctuations may become important however and may contribute to the dispersion in the racetrack array. To study its effect on particles we can use the nondimensionalized Langevin equation
\begin{equation}
    \dot x = F(x)+ f + \xi(t).
    \label{eq:langevin}
\end{equation}
where $F(x)$ is the same random force field as in Eq.(\ref{eq:monomereq}) and $\xi(t)$ is a standard Langevin noise with the properties $\overline{\xi(t)}=0$, and $\overline{\xi(t)\xi(t')}=2T\delta(t-t')$ averaging over the ensemble of possible noise realizations. Here $T$ the nondimensionalized temperature (to get the physical temperature we multiply by the unit of temperature $f_0 d_0/k_B$). For simplicity and to compare with that case, we will consider $F(x)$ to be of the same types discussed in section \ref{sec:boxdist}.

At variance with the other deterministic cases analyzed before, before discussing the dispersion properties of Eq.(\ref{eq:langevin}) it is important to remark here the distinction between the so-called ``annealed'' and ``quenched'' dispersion constants~\cite{LeDoussal1989,BouchaudGeorges1990}. 
{The ``quenched'' dispersion coefficient 
characterizes the spread of a packet in a single environment, while the ``annealed'' coefficient characterizes the spread of the configuration-averaged packet.
}
When we describe the normal spreading of a packet of particles in a single track due to finite temperature we are dealing with the ``quenched'' dispersion constant. From Fig.\ref{fig:dispersion} it is clear that we are instead interested in the so-called ``annealed'' dispersion constant $D$. In the absence of disorder, $v=f$ and $D=2T$, since it is standard driven Brownian motion, and then the two dispersion constant coincide. In the presence of disorder however the ``annealed'' is in general larger than the ``quenched'' dispersion constant~\cite{LeDoussal1989}.

In the presence of disorder and thermal fluctuations the problem becomes in principle quite complex. On one hand there is dispersion even below the depinning threshold, as the particles are able, in each track, to jump their energy barriers separating the metastable states by thermal activation. The stochastic dynamics is non trivial and can lead in general to anomalous dispersion, specially at low dimensions and for correlated random forces. We refer the reader to Ref.\cite{BouchaudGeorges1990} for a detailed review on this problem. 
Above the putative depinning threshold however, advection dominates and destroy 
any long-range time correlations. Also, the residence time in each cell has a finite mean and variance for any $T$. Hence, we expect the Central Limit Theorem to hold and normal dispersion to have a well defined $D$. Our aim here is to describe such regime and show how $D$ behaves vs $f$ and $T$, particularly well above the depinning threshold where the robust power-law scaling emerges as a function of $f$ or $v$.

In Fig.\ref{fig:temperature}(a)-(b) we show 
the effect of temperature on the quadratic mean displacement vs time for different temperatures in the RF and RB cases respectively. As can be observed, dispersion is normal and temperature monotonically increases dispersion. To better appreciate its effect 
in Fig.\ref{fig:temperature}(c) we show that the effect of temperature is approximately additive at large $f$: $D(f,T)\approx D(f,T=0)+ 2T$, regardless of the type of disorder. 

\subsubsection{Summary}
The effect of temperature for the driven overdamped particle in a RF or RB disorder is just to add the trivial thermal dispersion constant $D_0=2T$, describing dispersion in the absence of disorder. This result is expected as the disorder-induced fluctuations become small compared with the (fixed strength) thermal bath-induced fluctuations. 
Therefore, the scaling (Eqs.(\ref{eq:RFresult}) and (\ref{eq:RBresult})) remain robust in the sense that now $D(f,T)-D_0 \approx f^{-1} \approx v^{-1}$ for RF and $D(f,T)-D_0 \approx f^{-3} \approx v^{-3}$ for RB.

\subsection{A model for magnetic domain walls in wires}

The dynamics of magnetic domain walls in narrow tracks has attracted a great interest after the proposal of using it to devise a non-volatile racetrack memory~\cite{Parkin2008}.
As a starting point for modelling a magnetic domain wall (DW) driven by either an applied magnetic field and/or a current we will follow Ref.\cite{Thiaville2005} and consider the modified Landau-Lifshitz-Gilbert equation for the local magnetization vector $\vec{m}$,
\begin{eqnarray}
\dot{\vec{m}} &=& 
\gamma_0 \vec{H} \times \vec{m} + 
\alpha \vec{m} \times \dot{\vec{m}} -
(\vec{\mathcal{U}}.\vec{\nabla})\vec{m} 
\nonumber \\
&+& \beta \vec{m} \times 
[(\vec{\mathcal{U}}.\vec{\nabla})\vec{m}],
\label{eq:micromagnetics}
\end{eqnarray}
where $\vec{m}$ is a unit vector, $\gamma_0$ the gyromagnetic constant, 
$\vec{H}$ the micromagnetic effective field and $\alpha$ the Gilbert damping constant.
The velocity $\vec{\mathcal{U}}$ is a vector directed along the direction of electron motion, with an amplitude $\mathcal{U}=JPg\mu_B/(2eM_s)$, where  $M_s$ is the spontaneous magnetization, $J$ is the current density and $P$ its polarization rate. The non-dimensional adiabatic parameter $\beta$ is expected to be comparable to $\alpha$. Thermal fluctuations are neglected.

Solving Eq.(\ref{eq:micromagnetics}) for the dynamics of a driven domain wall is computationally expensive, due to the large number of degrees of freedom. Indeed, even in one dimension, where $\vec{m}\equiv \vec{m}(x,t)$, the system is extended. An usual approximation to leverage this difficulty is the collective coordinate approach which assumes a flat DW profile characterized by only two degrees of freedom: the DW position $q$ along the wire longitudinal $x$ direction, and the internal degree of freedom $\phi$ indicating the orientation of the magnetization at the DW center. For a transverse domain wall Thiaville \textit{et al.} \cite{Thiaville2005}
have obtained, for an homogeneous material,
\begin{eqnarray}
  \frac{\alpha}{\Delta}\dot q + \dot \phi &=& \gamma_0 (H_a+ H_p(q)) + \frac{\beta}{\Delta}\mathcal{U}  \\
  \alpha \dot \phi - \frac{\dot q}{\Delta} &=& -\frac{\gamma_0 H_K}{2} \sin 2\phi - \frac{\mathcal{U}}{\Delta}
\end{eqnarray}
where $\Delta=[A/(K_0 + K \sin^2 \phi)]^{1/2}$, with $K_0$ and $K$ the effective longitudinal and transverse anisotropy respectively, and $H_K=2K/\mu_0 M_s$. For simplicity we will make a rigid wall approximation or assume that $K \ll K_0$, so to approximate $\Delta$ with a constant. We will also assume that quenched disorder (arising from intrinsic magnetic or non-magnetic defects in the material, thickness modulations or from the wire borders roughness) gives place to an extra random field coupled only to $q$, $H_p(q)$, as in Lecomte \textit{et al.} \cite{Lecomte2009}.

By measuring $H_a$ and $H_p(q)$ in units of $H_K/2$, $u$ in units of $v_W=\Delta \gamma_0 H_K/2$,  time in units of $(1+\alpha^2)\Delta/v_W$, and $q$ in units of $\Delta$ we get (overriding notation for nondimensionalized $q$, $\phi$ and $t$), the nondimensional dynamical system  
\begin{eqnarray}
  \dot q  &=& \alpha [h + F(q)] + \sin(2\phi)+ u (1+\alpha \beta) 
  \label{eq:microm1}
  \\
  \dot \phi &=& h+ F(q) + (\beta-\alpha) u -\alpha \sin(2\phi),  
\label{eq:microm2}
\end{eqnarray}
controlled by the dimensionless parameters $\alpha$ and $\beta$, and drivings $h=2H_a/H_K$ and $u=\mathcal{U}/v_w$, with $F(q)=2H_p(q)/H_K$ the dimensionless pinning field.
It is worth noting that if we freeze the internal degree of freedom $\phi$, of the DW the model reduces to the overdamped mechanical model, essentially identical to Eq.(\ref{eq:monomereq}). 
We also note that in the absence of disorder ($F(q)=0$) at long times $\dot \phi=0$ if $h+(\beta-\alpha)u < \alpha$. This threshold corresponds to the model approximation of the Walker breakdown, separating the so-called stationary regime, where $\alpha \dot q = h + \beta u$ and $\dot \phi=0$, from the precessional regime where $\langle \dot \phi \rangle \neq 0$. With pinning, there is a fixed point $\dot q = \dot \phi =0$ or pinned state for small enough yet finite drives.
To solve Eqs.(\ref{eq:microm1}) and (\ref{eq:microm2}) we will consider again the piece-wise random-force field $F(q)$ used in section \ref{sec:boxdist}, both in the RF and RB cases. We will consider both the case of pure magnetic field driving ($u=0$) and the pure current driving $(h=0)$, and the cases $\langle \dot \phi \rangle=0$ and $\langle \dot \phi \rangle \neq 0$, 
and will focus on their large velocity regimes, $v\equiv \langle \dot q \rangle \sim u(1+\alpha \beta)$ 
and 
$v\equiv \langle \dot q \rangle \sim \alpha h$, respectively.

\subsubsection{RF case}
\begin{figure}[htp]
    \centering
\includegraphics[width=\columnwidth]{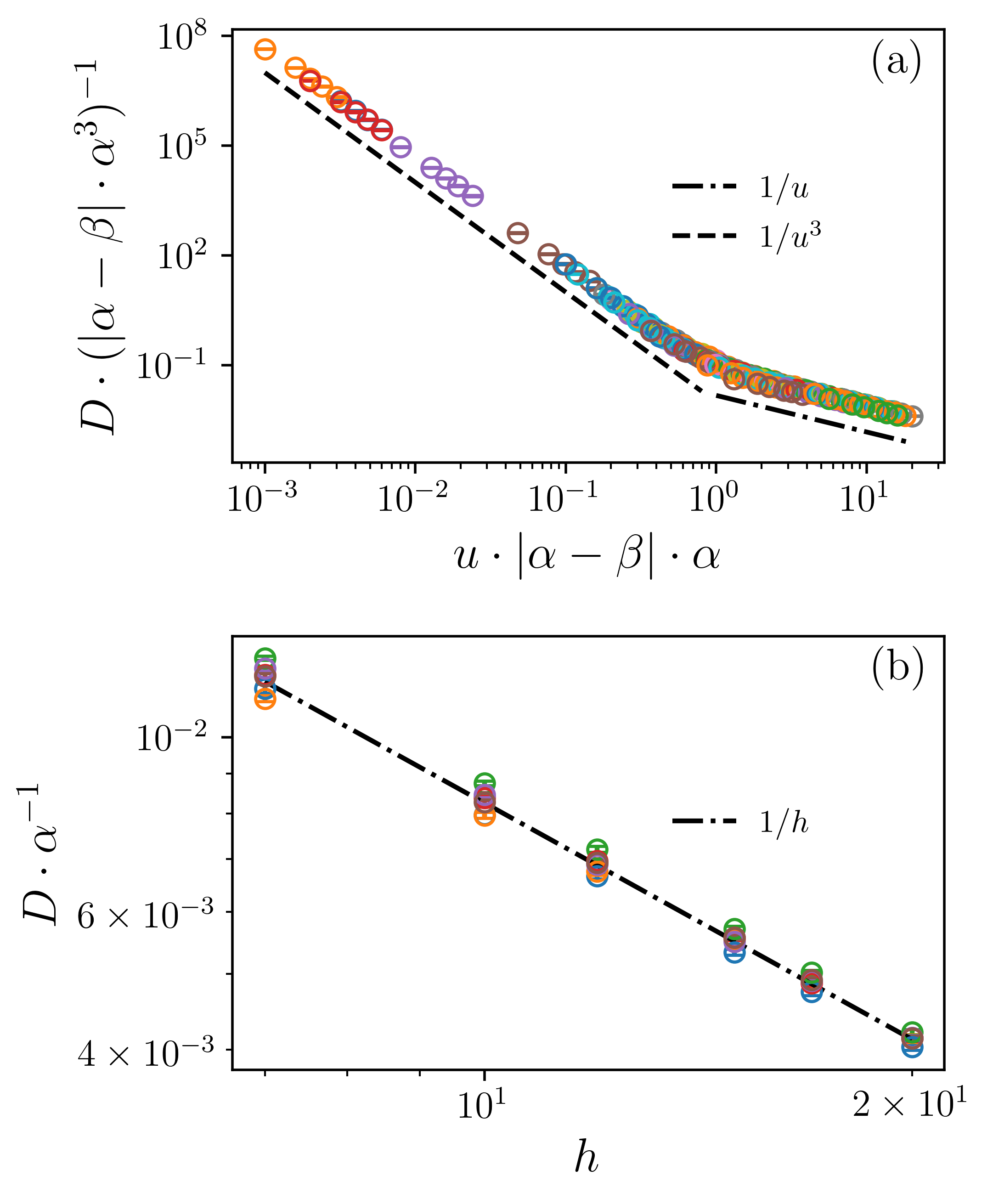}
    \caption{Scaled dispersion constant of a domain wall with RF disorder, from the numerical solution of Eq.(\ref{eq:microm2}). In (a) we show the current driven case ($h = 0,\; u > 0$), for $\alpha=\{0.02, 0.2, 0.25, 0.35 0.5, 0.75, 1.0\}$ and $\beta = \{0.0,0.02, 0.04, 0.08, 0.1, 0.2\}$ (all $\{\alpha, \beta\}$ pairs). The data is in agreement with a scaling of the form $D=|\alpha-\beta|\alpha^3 G_u(u|\alpha-\beta|\alpha)$ where $G_u(x) \sim x^{-1}$ for $x \gg 1$ and $G_u(x) \sim x^{-3}$ for $x \ll 1$. In (b) we show the pure field driven case ($u = 0,\; h > 0$) for $\alpha=\{0.01, 0.02, 0.2, 0.35 0.5, 1.0\}$. No crossover is observed and the data follows approximately the relation $D/\alpha \propto h^{-1}$.
    \label{fig:DWRF}}
\end{figure}

In Fig.\ref{fig:DWRF}(a)-(b) we show the results for the RF case for the current driven ($h=0$, $u>0$) and field driven ($h>0$, $u=0$) cases respectively, for drivings well above the corresponding depinning thresholds. 
In Fig.\ref{fig:DWRF}(a) we show the rescaled dispersion constant $D$ for the current driven case for $u$ values such that $v\equiv \langle \dot q \rangle \approx u(1+\alpha\beta)$. We empirically find that $D=|\alpha-\beta|\alpha^3 G_u(u|\alpha-\beta|\alpha)$, with $G_u(x)$ a scaling function, describes well the data for different pairs of values $(\alpha,\beta)$ which are sound for real materials. At large $u|\alpha-\beta|\alpha \gg 1$ we find $D\sim u^{-1}$ in agreement with the toy model with a RF disorder, Eq.(\ref{eq:RFresult}). Interestingly however, for smaller $u|\alpha-\beta|\alpha \ll 1$ a crossover towards a $D\sim  u^{-3}$ is observed, coinciding with the decay predicted from the toy model for a RB disorder.  
In Fig.\ref{fig:DWRF}(b), for the field driven case, we show that $D \approx h^{-1}$ at large $h$, with a power-law decay in $h$ in agreement with the RF case of the toy model, Eq.\ref{eq:RFresult}. Interestingly, no crossover is observed in this case.

A qualitative explanation of the results of Fig.\ref{fig:DWRF} can be given.
The crossover in the current driven case (Fig.\ref{fig:DWRF}(a)) occurs at $u^*\approx |\alpha-\beta|^{-1}\alpha^{-1}$. For $u>u^*$, we have $D\approx \alpha^2/u$, which is consistent with an RF disorder of amplitude $\alpha$ in the one degree of freedom toy model, Eq.(\ref{eq:RFresult}). Therefore, the $\sin(2\phi)$ term in Eq.(\ref{eq:microm1}) appears to have a negligible influence in the dispersion compared with the one expected from the term $\alpha F(q)$. To explain this we note, on one hand, that the larger $\alpha$ (the smaller $u^*$), the more important $\alpha F(q)$ is compared with the average effect of $\sin(2\phi)$ on the dispersion.
One the other hand, when $u<u^*$ dispersion is dominated by the $\sin(2\phi)$. Subtly, its effect on the  dispersion of $q$ is similar to the one of a RB disorder, yielding $D\sim 1/u^3$. Nevertheless, this case can not be described by a simple one degree of freedom toy model such as Eq.(\ref{eq:monomereq}) for an effective RB force field $F(q)$ (see appendix \ref{sec:reducido}).
The above arguments also explain qualitatively the field driven case of Fig.\ref{fig:DWRF}(b), since in this case the characteristic frequency of fluctuations of the $\sin(2\phi)$ term is always of order $h$ and can not be reduced by changing $\alpha$ and/or $\beta$. Since $v\sim h$, the average effect of $\sin(2\phi)$ can be neglected and then $D\sim \alpha^2 h^{-1}$ as expected for the toy model in a RF disorder of amplitude $\alpha$, Eq.(\ref{eq:RFresult}).

\subsubsection{RB case}
\begin{figure}[htp]
    \centering
\includegraphics[width=\columnwidth]{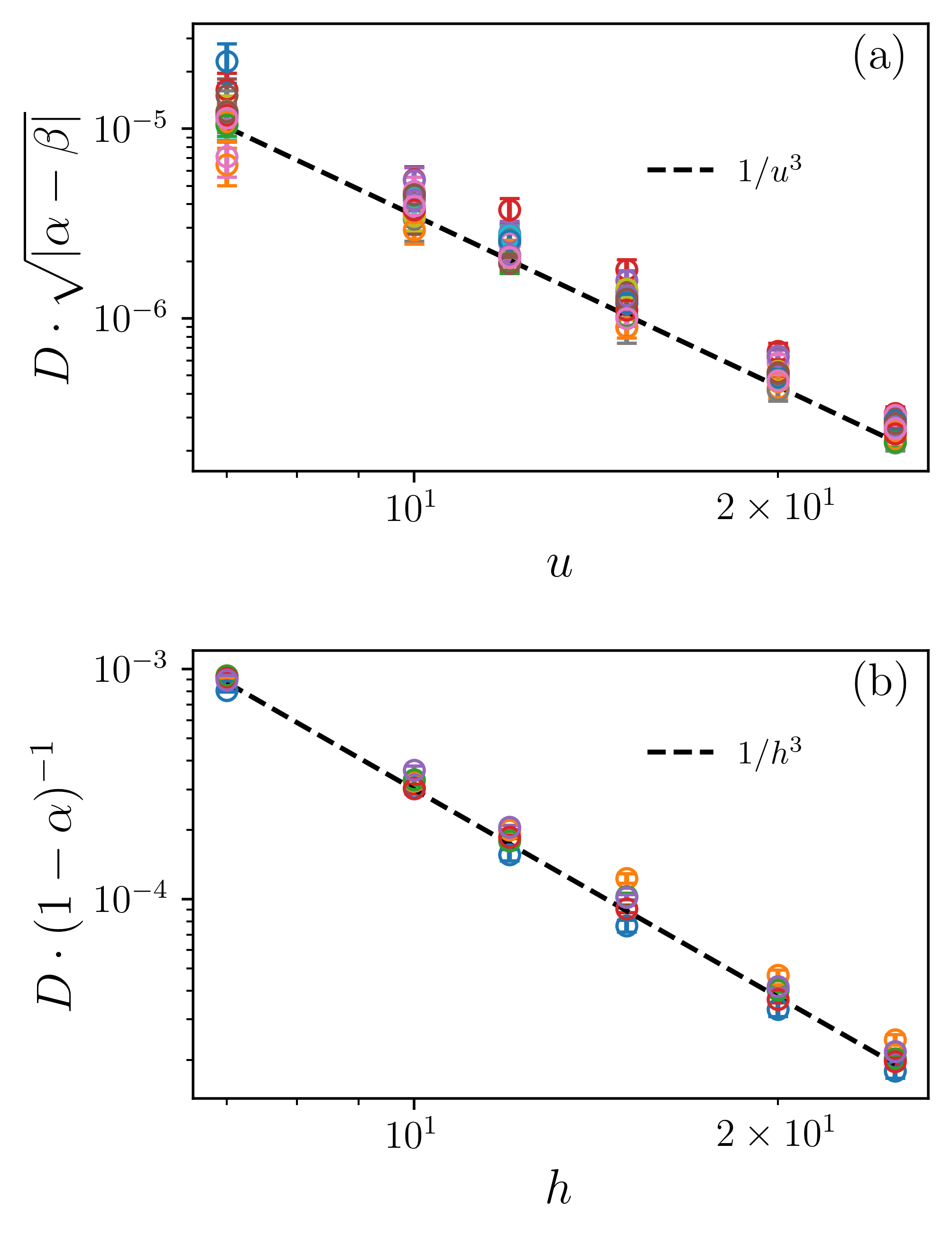}
    \caption{Scaled dispersion constant of a domain wall with RB disorder for $\alpha=\{0.3, 0.4, 0.5 0.6, 0.8\}$. In (a) the pure current driven case ($h = 0,\; u > 0$) for $\beta = \{0.0 , 0.02, 0.04\}$ shows the expected RB behavior as $D \cdot \alpha \propto 1/u^3$. In (b) the pure field driven case ($u = 0,\; h > 0$) shows again the expected RB behavior but $D \cdot (1-\alpha)^{-1} \propto 1/h^3$. In both cases, no crossover is observed, at variance with the current driven RF case (see Fig.\ref{fig:DWRF}(a)).}
    \label{fig:DWRB}
\end{figure}

In Fig.\ref{fig:DWRB}(a)-(b) we show the results for the RB case for the current driven ($h=0$, $u>0$) and field driven ($h>0$, $u=0$) cases respectively, from the numerical solution of Eq.(\ref{eq:microm2}). We observe $D\sim u^{-3}$ and $D\sim h^{-3}$, in agreement with the toy model prediction Eq.(\ref{eq:RBresult}) for RB disorder.
Interestingly, no crossover is observed in the current-driven case of Fig.\ref{fig:DWRF}(a). Using the arguments above, this may be explained by the fact that in this case both, $\alpha F(q)$ and $\sin(2\phi)$, are of a RB type, and thus $D\sim u^{-3}$ is in agreement the toy model, Eq.(\ref{eq:RBresult}) for RB disorder.
Similarly, the field driven case can be explained by the fact that $\sin(2\phi)$ in Eq.(\ref{eq:microm1}) oscillates at a frequency $\sim 2h$ larger than $v$, which is the characteristic frequency induced by $\alpha F(q)$. Therefore, its average effect on dispersion vanishes, and the system effectively behaves as the toy model in the RB case, (Eq.\ref{eq:RBresult}).
Power-law prefactors in Figs.\ref{fig:DWRB}(a)-(b) are different and less simple to predict in the RB case as compared to the RF case.
The scaling collapse in  Figs.\ref{fig:DWRB}(a)-(b) 
is purely empiric. 

\subsubsection{Summary}
In summary, we have shown that a zero-dimensional domain-wall derived from micromagnetic considerations, presents the following features:
\begin{itemize}
    \item At large drives, either from an applied current or magnetic field, $D$ decays as expected from the toy model predictions Eqs.(\ref{eq:RFresult})-(\ref{eq:RBresult}) for the RF and RB cases respectively.
    \item In all cases $D$ scales differently  with respect to the Gilbert damping parameter $\alpha$ and the adiabatic parameter $\beta$. 
    \item A crossover between different power-law decays is observed in $D$ only in the RF case, for the range of $\alpha$ and $\beta$ analyzed.
\end{itemize}

\section{Discussion}
\label{sec:discussion}
Our results show that the way the dispersion constant $D$ vanishes at large velocities is independent of many model details. It is nevertheless particularly sensitive to the nature of disorder. One may consider other types of random-force fields, such as long-ranged or periodically correlated ones, and the way $D$ vanishes may change. The RB and RF cases are nevertheless rather important physical cases to start with. For zero-dimensional magnetic domain walls for instance (i.e. when can neglect its transverse fluctuations and consider it as a flat interface), the RB type is generated by non magnetic-impurities, rough borders,  modulated thickness or by a space-dependent magnetic anisotropy, while the RF is realized in the presence of magnetic-impurities inducing a random magnetic field. The situation is similar for zero-dimensional ferroelectric domain walls, where the two types of quenched disorders can be in principle realized. On the other hand, magnetic skyrmions, colloids or vortices typically couple to a random bounded potential and hence realize a RB disorder.

Extended systems may be also considered, such as one- or two-dimensional domain walls anchored by a narrow channel of a finite width. Skyrmions or soliton chains in one dimensional tracks \cite{Osorio2022} also fall in the extended system category. In these systems with many elastically coupled degrees of freedom the random-force acting on the center of mass may be renormalized. A prominent example are elastic walls in a short-ranged correlated random-force field, for which a RB microscopic disorder is renormalized to a cuspy RF-type near depinning. The force-force correlator then gets rounded at larger velocities as the dynamical correlation length reduces~\cite{chauve2000,Wiese2022}. In these cases we expect our results for the dispersion to hold for the center of mass respecting the shape of the renormalized disorder instead of the microscopic one. It is plausible that, in such case, a scenario similar to the one described in section \ref{sec:rb+rf} takes place at large velocities. Indeed, for a driven elastic vortex line in three dimensions, with RB disorder, an effective temperature vanishing as $\sim v^{-3}$ was found  for a wide range of velocity in spite that the depinning transition is better described by a renormalized RF-type random force field~\cite{Elias2022}. 
It is worth also noting that in the context of mean-field theory the drive on the effective one particle models is a moving parabolic well instead of a constant force, leading to a bounded dispersion. Nevertheless, the force-force correlator,  both at zero~\cite{LeDoussal2009,terBurg2021} and finite-temperature~\cite{terBurg2022} remains a key common property of both ensembles. It would be interesting to connect the properties of these force-force correlators with the center of mass dispersion of extended objects in a racetrack array.

Bringing back the dimensions, our results show that at large velocities, we can roughly expect 
\begin{eqnarray}
v&\approx& (v_0/f_0)f, \\
D&\approx& v_0 d_0 (f_0/f)^n,   
\end{eqnarray}
where $f_0$ is the pinning force amplitude, $d_0$ the disorder correlation length  $v_0=\eta f_0$ is the characteristic velocity, $\eta$ the friction constant, and $n=1$ or $n=3$ for RF and RB disorder respectively. To compare with experiments we may use that the experimentally accessible depinning force is $f_c \sim f_0$ (neglecting interaction effects between particles) and hence $v_c \equiv \eta f_c \sim v_0$.  The disorder correlation length $d_0$ is usually unknown or difficult to estimate directly, but one can use that $d_0=\max(R,r_0)$, where $R$ is the size of the particle and $r_0$ the correlation length-scale of the inhomogeneity. We can thus write a more practical and simple prediction
\begin{eqnarray}
    D\approx v d_0 (f_c/f)^{n+1} \approx v d_0 (v_c/v)^{n+1}.
    \label{eq:Dwithunits}
\end{eqnarray}
From an applied viewpoint, to realize a memory bus analog, it would be desirable to have particles in different tracks concurrently and precisely controlled. If we need for instance to transport coherently and at the largest possible velocity a few bytes from one check point to the next by moving particles in different parallel tracks, the less possible dispersion would be ideal. To make a concrete estimation let us define the maximum time $t_{\rm max}$ and distance $x_{\rm max}$ we can move the particles at given velocity $v$  with a given tolerance $\delta$ for the dispersed packet width, i.e $D t_{\rm max}= \delta^2$. It makes sense to ask for a maximum tolerated error to be of the order of one particle (one bit) size, 
i.e. $\delta \sim R$.
We then get
\begin{eqnarray}
  x_{\rm max} \equiv 
  v t_{\rm max} \approx 
  \frac{R^2}{d_0}\left(\frac{f}{f_c}\right)^{n+1} \approx 
  \frac{R^2}{d_0}\left(\frac{v}{v_c}\right)^{n+1}.
  \label{eq:predictions}
\end{eqnarray}
For skyrmions for instance, if we assume point disorder, we have $d_0 \approx R$. Then, using a skyrmion size $R\approx 10 \text{ nm}$ moving at $v\approx 0.34 \text{ mm/s}$ for $f/f_c \equiv j/j_c \approx 5$~\cite{Fert2013} we obtain $x_{\rm max} \approx 6 \text{ }\mu\text{m}$ and $t_{\rm max}\approx 0.025\text{ s}$ for a coherent parallel transport. If the disorder were otherwise of a pure RF type ($n=1$) then $x_{\rm max}$ and $t_{\rm max}$ would be more than one order of magnitude smaller. On the other hand, if the microscopic correlation length of the inhomogeneity is $r_0>R$ the previous estimates for $x_{\rm max}$ and $t_{\rm max}$ enhance by a factor $R/r_0$. 
Rough estimates similar to the above can be made for domain walls, superconducting vortices or colloids, provided we know a few parameters such as the size of the particles 
and the velocity-force characteristics under the assumption of a short-ranged random force correlator.  Besides specific applications however, Eq.(\ref{eq:Dwithunits}) is a rather robust prediction which might be useful to characterize the disorder in the host media, regardless of its microscopic origin, by measuring the dispersion $D$ of independent racing particles as a function of the driving force. 
\begin{acknowledgments}
We thank Kay Wiese for illuminating discussions. We also acknowledge support from grant PICT-2019-01991.
\end{acknowledgments}

\appendix
\section{Dispersion constant in the piece-wise quenched random force field}
\label{sec:appendixformulaD}
The mean velocity $v$ is such that in the long time limit
$\langle [x(t)-v t] \rangle=0$. If $m \equiv \langle \Delta t \rangle$ is the mean time in a unit cell, then the mean velocity is $v=1/m$. In order to calculate the quadratic mean displacement 
$\langle [x(t)-v t]^2 \rangle$ as a function of time in the nondimensionalized model we can fix the distance as $N$ and consider the time $t_N$ needed to reach it as a random variable, i.e. $x(t_N)=N$. Then, considering $x(0)=0$ we have
\begin{eqnarray}
\langle [x(t_N)-v t_N]^2 \rangle &=& 
N^2 + v^2 \langle t_N^2 \rangle - 2 N v \langle t_N \rangle \nonumber \\
&=&
v^2 \langle t_N^2 \rangle - N^2
\end{eqnarray}
where we have used that $v\langle t_N \rangle=N$.
Let us calculate 
\begin{eqnarray}
\langle t_N^2 \rangle &=& 
\langle \sum_{n,m} \Delta t_n  \Delta t_m \rangle
=
\sum_{n,m} [\langle (\Delta t_n-m)( \Delta t_m-m) \rangle
+m^2] \nonumber \\
&=& 
\sum_{n,m} [\sigma^2 \delta_{n,m}
+\langle\Delta t\rangle^2] 
= N\sigma^2 
+N^2 m^2
\end{eqnarray}
where we defined $\sigma^2 \equiv \langle (\Delta t_n-\langle\Delta t\rangle)( \Delta t_m-\langle\Delta t\rangle) \rangle$ 
and we have used that residence-time fluctuations of different cells are independent.
Therefore 
\begin{eqnarray}
\langle [x(t_N)-v t_N]^2 \rangle &=& 
v^2 (N\sigma^2 
+N^2 m^2) - N^2 
\nonumber \\
&=&
v^2 N \sigma^2 = (v^2 \sigma^2 /m ) t. 
\end{eqnarray}
We can then identify the dispersion constant 
\begin{eqnarray}
D = v^2 \sigma^2 /m = \sigma^2/m^3
\end{eqnarray}
which is then completely determined by the mean and variance of $\Delta t$.
Since the Central Limit Theorem holds for 
the total displacement, the dispersion front at long times and for a large number of tracks, is
\begin{equation}
    P(x,t) = \frac{e^{-\frac{(x-v t)^2}{D t}}}{\sqrt{2\pi D t}}
\end{equation}
where $D$ and $v$ are both functions of $f$.\\

\section{Random friction model}
\label{sec:randomfriction}
A trivial analytically solvable model that displays a dispersion constant $D$ that increases with the driving force $f$ is
\begin{equation}
    (1+\gamma F(x)) \dot{x} = f.
\end{equation}
This equation of motion describes an overdamped particle motion 
with a friction force proportional to the instantaneous velocity, with a position dependent friction constant
$1+\gamma F(x)$. We will assume that $\gamma < 1$ (to assure a positive friction constant) and that $F(x)$ is a piece-wise random variable as the random force in section \ref{sec:boxdist}.
Therefore, the time $\Delta t$ spent while $x \in [n,n+1]$ is
\begin{equation}
    \Delta t = \frac{(1+\gamma F(x))}{f}. 
\end{equation}
Therefore, the average over tracks (or over disorder) yields
\begin{eqnarray}
    \langle \Delta t \rangle = \frac{1}{f},\;\;
    \langle \Delta t^2 \rangle -\langle \Delta t \rangle^2 = \frac{\gamma^2}{f^2}.
\end{eqnarray}
Hence, using Eqs.(\ref{eq:vformula}) and (\ref{eq:Dformula}) we have
\begin{eqnarray}
    v &=& f \\
    D &=& \gamma^2 f. 
\end{eqnarray}
This deterministic disordered model has no depinning transition and the annealed dispersion constant is proportional to $f$, at variance with the random-force models we presented in the main text.

\section{Reduced Model for current driven Domain Walls at small $\alpha$}
\label{sec:reducido}
\begin{figure}[h!]
    \centering
\includegraphics[width=\columnwidth]{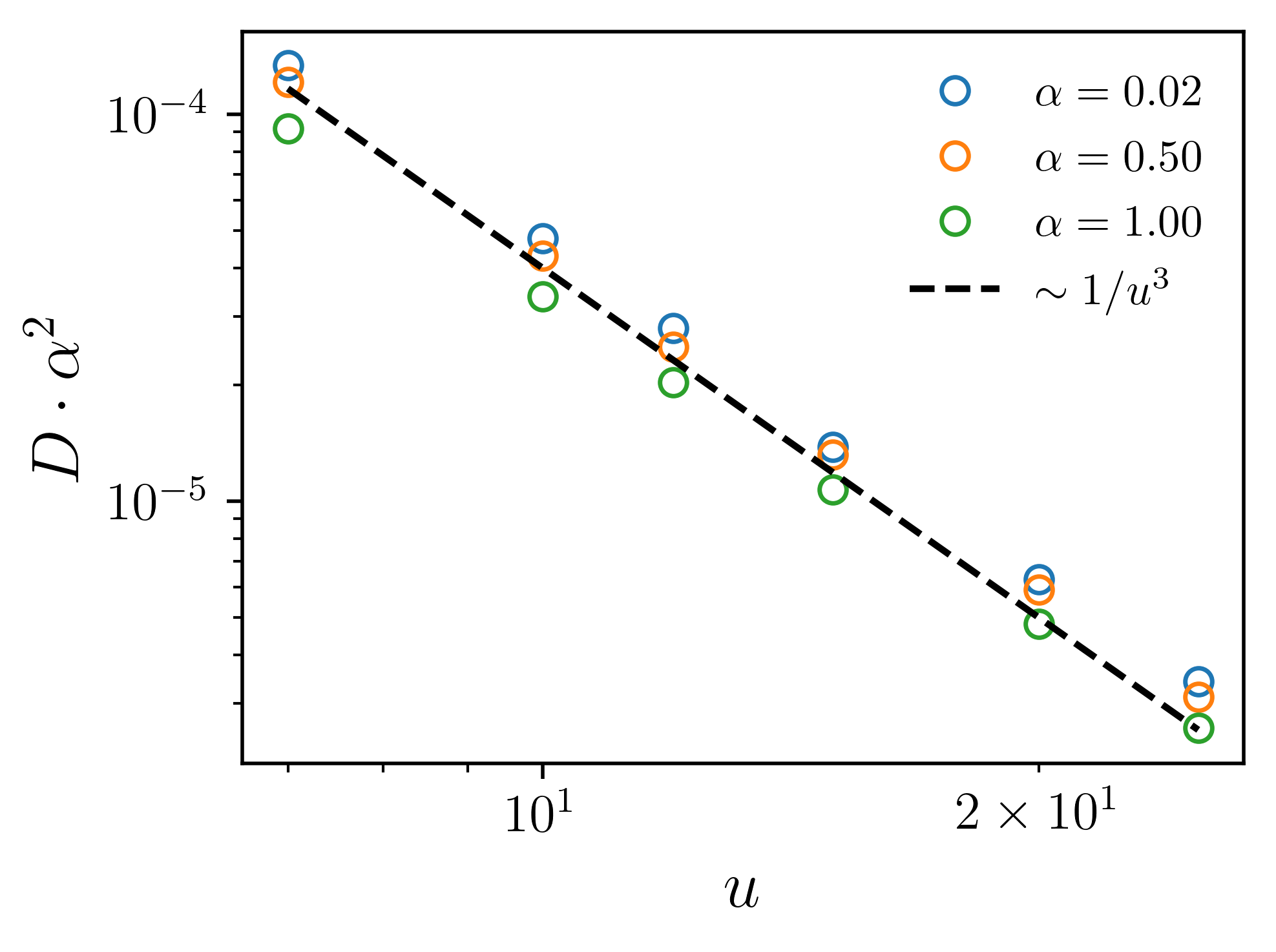}
    \caption{Scaled dispersion constant for the reduced model of Eq.(\ref{eq:reducedDW}) as a function of the current. The model shows,  for a RF $F(q)$, a $1/u^3$ decay as indicated with the dashed line.}
    \label{fig:reducido}
\end{figure}

A minimal model capturing the $D \sim 1/u^3$ behavior for the current-driven domain wall with RF disorder is obtained taking the limit small $\alpha$. In this limit the Eqs.(\ref{eq:microm1}) and (\ref{eq:microm2}) take the form
\begin{eqnarray}
    \dot q &=& \sin{2\phi} + u,\;\;\; 
    \dot \phi  = F(q)+\alpha u
        \label{eq:reducedDW}
\end{eqnarray}
where the disorder effect on the position is present indirectly, through the phase $\phi$. An equivalent picture is given by the integral equation
\begin{equation}
    \dot q = \sin{\left(\int_{0}^t dt'F(q(t'))+\alpha u t \right)} + u,
        \label{eq:reducedDW2}
\end{equation}
In Fig.\ref{fig:reducido} we show the dispersion constant for Eq.(\ref{eq:reducedDW})  with an RF random force $F(q)$. We observe $D \sim 1/u^3$ behavior, as indicated by a black dashed line. Although this is reminiscent to the RB induced dispersion of Eq.(\ref{eq:RBresult}), Eq.(\ref{eq:reducedDW2})
can not be reduced to a one degree of freedom in an effective RB short-ranged correlated random force field, such as the model of Eq.(\ref{eq:monomereq}),
suggesting a completely different and more complex mechanism.

\bibliography{biblio}{}

\end{document}